\documentclass[a4paper,11pt]{article}
\usepackage[a4paper, margin=1.2in]{geometry}

\usepackage{threeparttable} 
\usepackage{rotating}
\usepackage{float}          
\usepackage{graphicx}
\usepackage[colorlinks=true,
            linkcolor=blue,
            citecolor=blue,
            urlcolor=blue]{hyperref}
\usepackage{array}
\usepackage[utf8]{inputenc}
\usepackage{pdfpages}
\usepackage{rotating}
\usepackage{longtable}
\usepackage{booktabs,caption}
\usepackage{lscape}
\usepackage{blindtext}
\usepackage[english]{babel}
\usepackage{hanging}
\usepackage{dirtytalk}
\usepackage{authblk}
\usepackage{amssymb}
\usepackage{amsmath}
\usepackage{txfonts}
\usepackage{mathdots}
\usepackage[classicReIm]{kpfonts}
\usepackage{appendix}
\usepackage{float}
\usepackage{comment}
\usepackage[round]{natbib}
\usepackage{footmisc}

\newcommand{\sym}[1]{\ifmmode^{#1}\else\(^{#1}\)\fi}

\title{Beyond Time:\\ Unveiling the Invisible Burden of Mental Load}
\author[a]{Francesca Barigozzi}
\author[a]{Pietro Biroli}
\author[a]{Chiara Monfardini}
\author[a]{Natalia Montinari}
\author[a]{Elena Pisanelli}
\author[a]{Sveva Vitellozzi}
\affil[a]{\small \textit{Dept. of Economics, University of Bologna (Italy)}}

\begin{document}
\maketitle

\date{}

\vspace{-1.1cm}

\begin{abstract}
\noindent 
This paper introduces a novel, scalable methodology to measure individual perceptions of gaps in mental load—the cognitive and emotional burden associated with \textit{organizing} household and childcare tasks—within heterosexual couples. Using original data from the TIMES Observatory in Italy, the study combines time-use diaries with new survey indicators to quantify cognitive labor, emotional fatigue, and the spillover of mental load into the workplace. Results reveal systematic gender asymmetries: women are significantly more likely than men to bear organizational responsibility for domestic tasks, report lower satisfaction with this division, and experience higher emotional fatigue. These burdens are underestimated by their partners. The effects are particularly pronounced among  college-educated and employed women, who also report greater spillovers of family responsibilities than men during paid work hours. The perceived responsibility for managing family activities is more strongly associated with within-couple gaps in time use than with the absolute time spent on their execution, underscoring the relational and conflictual nature of mental load. 

\vspace{.4cm}
\end{abstract}
\footnotetext{\scriptsize \textit{Corresponding author}: Sveva Vitellozzi, e-mail: \href{mailto:sveva.vitellozzi@unibo.it}{sveva.vitellozzi@unibo.it}.\\
\textit{Acknowledgements and Funding}: This study was funded by the European Union – NextGenerationEU, in the framework of the “GRINS - Growing Resilient, INclusive and Sustainable project” (PNRR - M4C2 - I1.3 - PE00000018 – CUP J33C22002910001). The views and opinions expressed are solely those of the authors and do not necessarily reflect those of the European Union, nor can the European Union be held responsible for them. We are grateful to Margherita Fort for her suggestions and insights. We also thank Roberto Nisticò, Valeria Zurla, and  the participants of the online workshop organized by GRINS on February 21, 2025, for their valuable comments and suggestions. Matilde Penna and Hanieh Hassanshahi provided excellent research assistance.}


\section{Introduction}

Although managing a household cannot be equated with managing a business, it similarly demands substantial mental resources and time. However, unlike a manager who typically maintains some emotional distance from the enterprise, the person responsible for a household is often deeply emotionally invested, which intensifies the strain and fatigue associated with the role.\par
Managing a household entails a set of responsibilities often referred to as the “mental load,” which includes both \textit{cognitive labor}—the planning, organizing, and coordination of everyday family tasks—and \textit{emotional labor}—the sense of responsibility for the well-being of others within the household \citep{dean2022mental}. This paper contributes to the limited but growing body of research showing that mental load is another crucial (yet overlooked) dimension of gender inequality.\par
In recent years, time-use diary studies have significantly deepened our understanding of household dynamics, particularly with respect to gender gaps in time use. This body of research has consistently documented that women devote substantially more time than men to unpaid work, both household chores and informal childcare—even within dual-earner couples and in the most equalitarian countries.\footnote{See, among many others, \cite{mencarini2004time};  \cite{anxo_gender_2011}; \cite{gimenez2012trends}; \cite{barigozzi2023gender}.} Because it is not easily bounded in time or space—often occurring concurrently with other tasks and running in the background of daily life—mental load remains largely invisible in conventional time-use data \citep{offer2011revisiting, reich2023gendered}. Yet, evidence exists that it is disproportionately borne by women and is linked to stress, fatigue, lower productivity, lower well-being, and reduced relationship satisfaction \citep{offer2014costs, ciciolla2019invisible,aviv2025cognitive, vitellozzi2025invisible}. \par
In this paper, we propose a novel, scalable measurement approach to mental load and integrate it into a new survey which for the first time combines mental load measures with detailed time use data collected through time diaries. Leveraging data from 415 heterosexual couples in Italy, we offer new evidence on the gendered distribution of mental load and its correlates in everyday life and the workplace. The richness of our data allows us to investigate mental load along three complementary dimensions together with its relationship to time use allocation within the couple: (i) cognitive labor, proxied by who in the couple is perceived to be the main organizer of household and childcare activities; (ii) emotional load, captured by self-reported fatigue for family well-being; and (iii) spillover into the workplace, based on self-reports of thinking about domestic responsibilities while at work.\par
We begin by focusing on the cognitive dimension of mental load, examining the gendered distribution of two types of organizing responsibilities. Specifically, we consider: (a) the management of household tasks (e.g., deciding what to cook, which groceries to buy, overseeing cleaning, etc.); and (b) the organization of childcare activities (e.g., scheduling pediatric appointments, arranging after-school activities, etc.). We then examine how responsibility for these organizational tasks relates both to the time each partner devotes to carrying them out in daily family life and to the gender gap in time use at the couple level. We then examine how the perception of being the main organizer is linked to individuals’ satisfaction with the division of these responsibilities. Next, we turn to the \textit{emotional dimension }of mental load, analyzing how these perceptions relate to emotional strain, and whether unequal divisions of time devoted to the execution of household and childcare tasks between partners help explain feelings of mental burden. Finally, for employed respondents, we explore how cognitive labor spills over into paid work, investigating whether individuals continue to mentally manage the household during working hours.\par
Our results reveal striking gender asymmetries in the experience of mental load. Women overwhelmingly assume responsibility for organizing household and childcare activities: 63\% of women report being the primary organizer, and 52\% of men confirm that their partner holds this role. However, this organizational burden does not translate into higher satisfaction. On the contrary, women—especially those who are employed—report lower satisfaction with the division of organizing responsibilities. Men, by contrast, tend to express satisfaction when their partner manages the organization of household tasks, but are less satisfied when their partner is in charge of organizing children’s activities. Importantly, we find that the division of organizational responsibilities for household and childcare activities is closely tied to the actual time spent executing those same activities. Specifically, women who organize more of a given activity also tend to perform relatively more of that activity, underscoring how managerial and operational tasks are often intertwined in the domestic sphere. This burden continues beyond the home: among employed individuals, women are significantly more likely than men to report thinking “very often” or “always” about the organization of family responsibilities during paid work, pointing to a clear spillover of cognitive load into the workplace.\par
Our results suggest that \textit{emotional load} also falls disproportionately on women because female respondents report greater feelings of responsibility for their family’s well-being and higher levels of fatigue associated with this role. These emotional costs are especially reported by college-educated women, suggesting that higher education—
possibly linked to more egalitarian beliefs or higher expectations—intensifies sensitivity to unequal domestic arrangements. In our analysis, the strongest predictor of perceived emotional fatigue is dissatisfaction with the division of organizational tasks, while the strongest predictor of unequal sharing of managing responsibilities is the time-use gap between partners in the execution of those activities. These findings suggest that mental load—both in its cognitive and emotional dimensions—is a gendered burden and one that extends into the workplace: employed women are more likely to report thinking about household management during work hours, highlighting the spillover effects of this invisible labor.\par

Contrary to common narratives suggesting that women derive fulfillment from organizing household activities, our results suggests that this responsibility is associated with dissatisfaction and fatigue. Mental load emerges as a unequally distributed burden, particularly among more educated women.\par
 While prior studies have offered conceptual and qualitative insights (e.g., \cite{offer2014costs, daminger2019cognitive}) or focused on maternal well-being (\cite{ciciolla2019invisible, dean2022mental}), few have provided tools for systematic measurement. 
Recent work in psychology has developed validated scales to assess the “invisible family load” using mixed-methods designs \citep{wayne2023s}. However, these instruments remain limited to capturing individual perceptions and are not designed to link systematically with partner-level time-use data, limiting their applicability in representative household studies. Our paper proposes a novel and scalable approach to measuring mental load, that can be easily integrated into time-use surveys allowing for a more exhaustive picture of gender inequalities within the couple. In Section \ref{sec:related.liteature}, we outline how our contribution advances the existing literature.

\section{Data}
\subsection{TIMES data}
This study relies on data from the TIMES Observatory (\href{https://site.unibo.it/times/it}{Osservatorio sull’Uso del Tempo nelle Giovani Famiglie}), a new and ongoing data collection project developed by researchers at the University of Bologna, within the broader GRINS (Growing, Resilient, Inclusive, and Sustainable) partnership. The TIMES project is designed to investigate time use, paid and unpaid labor, and family dynamics in households with children, with a strong focus on gender roles and intra-household responsibilities.\par
The sample is composed of couples with children under the age of 11.\footnote{The sample includes cohabiting couples (regardless of gender) with at least one child under the age of 11 living in the household. While the broader dataset (N = 1124) allowed for up to 25\% of respondents to participate individually—i.e., without their partner—this study focuses exclusively on couples in which both partners completed the survey (N = 832). One same-sex couple is excluded from the analysis due to limited representation in the sample. Sampling was stratified by province, gender, employment status, and urbanity of residence, and also accounted for the distribution of children by age group and geography.}
It was recruited ad hoc by a professional survey research company with extensive experience in large-scale data collection. 
Participants were sampled to be representative of the population of the Emilia-Romagna region, stratified by province, gender, employment status, and urbanity of residence. Quotas also accounted for the distribution of children under 11 by age group and geography.
Data collection was conducted over twelve months (December 2023 to December 2024), with monthly quotas to ensure temporal balance. Respondents completed a socio-demographic questionnaire and two time-use diaries (one weekday, one weekend day), using a web-based app developed specifically for the project. Diary days were randomly assigned and matched across partners, following a pre-specified randomization protocol.\footnote{Diary completion occurred within three days of the assigned reference date. Respondents were contacted via email, WhatsApp, or SMS, and received a financial incentive of €20 for full individual participation (questionnaire + diaries), or €50 if both partners in the couple completed the full survey.}

A key feature of the TIMES survey is that it collects data from both partners within the same household, enabling unique insights into the joint dynamics of time allocation, responsibility for household and care work, and perceptions of fairness and fatigue. To measure mental load we included specific questions informed by the most recent literature on mental load (more detailed information on mental load variables can be found below). To our knowledge, this is the first time that time diary data have been collected together with mental load measures. The data collection combines a detailed questionnaire with a time-use diary, employing a CAWI (Computer-Assisted Web Interviewing) methodology via a custom-developed web app. The web app is accessible via computer, tablet, or smartphone, and is available in both Italian and English.\par
The survey proceeds in two stages. In the first stage, participants respond to a socio-economic questionnaire covering--- besides our new mental load questions---demographics, work arrangements, division of labor, attitudes toward gender norms, fertility intentions, and parental leave use. In the second stage, approximately two weeks later, participants complete a time-use diary for two days (one weekday and one weekend day). This diary captures primary and secondary activities, as well as who is present and involved in each activity. This structure allows us to measure the intensity and distribution of time spent in paid work, unpaid domestic and care work, leisure, and multitasking. 

\par
Our new measures of mental load contained in TIMES data are particularly well suited to studying mental load as an invisible dimension of unpaid work. First, they allow us to understand the division of organizational responsibility within the household. Second, they capture self-reported emotional outcomes---such as feelings of fatigue related to household and caregiving responsibilities---allowing us to explore how individuals experience the burden of these responsibilities, which time use diaries cannot measure. Finally, by pairing perceptions of responsibility and fatigue with actual time-use information, TIMES data enables a descriptive analysis of the interplay between the cognitive and emotional burdens related to managing domestic and children's activities and the physical time spent on them (i.e., their execution).

\subsection{Mental load in TIMES data}

Mental load is a multidimensional construct encompassing both the cognitive and emotional effort required to manage household and care responsibilities. In designing the TIMES survey, we drew on recent conceptual developments in the literature to inform our approach to measuring this largely invisible form of labor. Because the experience of mental load is inherently subjective, our analysis centers on individuals’ own perceptions of organizational responsibility and emotional strain. These perceptions can vary significantly both across and within households. This focus on subjective experience is not only a methodological choice but a defining feature of our contribution: it allows us to capture how mental load is lived, felt, and distributed as perceived by individuals themselves.
\par
First, our decision to focus on both cognitive and emotional aspects of household labor is consistent with \cite{dean2022mental}, who argue that the defining feature of mental load is the combination of “thinking work” (planning, managing, anticipating) and its emotional weight. Similarly, \cite{daminger2019cognitive} identifies anticipation, planning, and monitoring as core components of cognitive labor, highlighting how these often precede or accompany the physical execution of tasks. These insights guided our inclusion of survey items that explicitly distinguish between who takes responsibility for organizing daily activities (e.g., who thinks about and plans them), and how individuals feel about this distribution.\par
Second, the literature underscores the gendered and affective nature of mental load. \cite{reich2023gendered}, in their systematic review, emphasize the importance of measuring not only task division but also the psychological cost associated with managing household roles—particularly among women. To capture this emotional dimension, we included items that ask respondents how much responsibility and fatigue they feel for different domains of family well-being (e.g., children, partner, household, family as a whole).\par
As a result, we focus on two key domains:
\begin{itemize}
    \item \textbf{Cognitive labor:} the mental effort involved in planning and organizing domestic and childcare activities.   
   
    \item \textbf{Emotional labor:} the psychological effort associated with feeling responsible
for others’ well-being and the emotional toll this entails. 

\end{itemize}

\subsubsection{Measuring Mental Load}
To measure mental load within the TIMES dataset, we rely on a set of self-reported indicators designed to capture both cognitive and emotional labor. Appendix B reproduces the English translation of the survey questions used to elicit the two dimensions of mental load.  \par
\bigskip
\textbf{Cognitive labor} is captured via questions on who organizes household and childcare activities, satisfaction with the division of this organization, and—among employed respondents—how often individuals think about these tasks during their workday.

    A caveat is necessary here. As previously noted, mental load is difficult to delimit in time and space, as it often occurs simultaneously with other activities and operates in the background of daily life. Consequently, it is challenging to measure directly using time diaries. 
    

    Our approach addresses this limitation by eliciting cognitive load indirectly—through self-reports on who holds primary responsibility for organizing household and childcare tasks, and how satisfied individuals are with this division. Rather than aiming to objectively quantify the burden of cognitive labor, we intentionally focus on perceived responsibility and emotional strain as experienced by individuals. This emphasis on subjective perception is a core strength of our methodology, allowing us to uncover the gendered dynamics of mental load as they are lived and internalized, even when not externally visible or jointly acknowledged within the couple.
    
We capture cognitive load through three sets of variables:
\begin{enumerate}
    \item \textbf{Division of organizational responsibility:} 
    Respondents are asked who in their household is primarily responsible for organizing household activities and children’s activities, using a five-category scale ranging from “exclusively me” to “exclusively my partner.” 
    

Because cognitive load is predominantly borne by women \citep{daminger2019cognitive}, we will examine the dimensions of time use and family life associated with \textit{women being identified as the main organizers of household and childcare activities}. In addition, we will estimate separate regressions for women and men. Hence, our analysis focuses on gendered asymmetries in cognitive load while accounting for both self-perceived and partner-attributed responsibility.\footnote{\label{fn:myfootnote}Our analysis focuses on individual perceptions of gender gaps in mental load, rather than on agreement within the couple about who is primarily responsible for organizational tasks. Specifically: (\textit{i}) in the women’s sample, the dependent variable equals 1 if the respondent reports being “exclusively” or “mostly” responsible for the organizational task; and (\textit{ii}) in the men’s sample, the dependent variable equals 1 if the respondent reports that his partner is “exclusively” or “mostly” responsible. As a result, consistency between partners’ responses is irrelevant for our indirect measure of cognitive load. However, since consistency of partners' responses is interesting \textit{per se} and informative about partners' overall awareness of mental load, we present summary statistics on within-couple response consistency below Figure \ref{fig:CL}, in this section, and in Appendix A.
}


    \item \textbf{Satisfaction with the division of organizational tasks:} respondents rate their satisfaction with the current division of organizational tasks for unpaid work--household and children's activities---on a scale from 0 (not at all satisfied) to 100 (fully satisfied). These variables allow us to assess perceived fairness and burden in the allocation of cognitive labor.
\item \textbf{Thinking about household organization at work:} among employed respondents, we include a question that asks how often they think about the organization of household and childcare tasks during their workday. Answers range from “never” to “always” on a 5-point scale. These items provide insight into how cognitive labor extends beyond the home and intrudes on paid work time. 

\end{enumerate}

\bigskip

\textbf{Emotional labor} is captured through subjective feelings of responsibility and fatigue related to various spheres of family life. Respondents are asked to rate, on a scale from 0 to 100, first how much \textit{responsibility} and, then, \textit{fatigue} they feel for:
\begin{itemize}
    \item their children’s well-being;
    \item their partner's well-being; 
    \item the execution of daily activities; and 
    \item the well-being of the family as a whole
\end{itemize}
These items allow us to capture the emotional costs associated with mental labor, particularly when it is perceived as burdensome. \par

While responsibility and fatigue are both important aspects of emotional labor, our empirical focus is on fatigue, which more directly captures the emotional cost of mental load. The underlying assumption is that feeling responsible does not necessarily constitute emotional burden; rather, emotional load arises when this sense of responsibility is experienced as fatiguing or psychologically taxing. In other words, it is not the act of caring or being responsible per se, but the emotional toll of managing this role that defines emotional labor.\par
Our main emotional load indicator consists of the mean score across the above-mentioned four fatigue-related items as a summary measure of emotional load. A higher average score on the indicator indicates a greater emotional burden associated with the management of family life. This measure aligns with theoretical work highlighting how the emotional component of mental load is shaped not only by what individuals do but also by how they internalize and experience their responsibilities.
\subsection{Time use variables}
The time use survey of the TIMES project includes two time-use diaries per respondent: one referring to a weekday and one to a weekend day. Both days were randomly assigned.
Respondents had to register, through a web interface, all the activities performed in the whole 24 h spectrum of the assigned day, selecting them from pre-determined lists. The design of the time diary was based on the work by \citep{bigoniforthcom}.  
Availability of information on two types of days - quite rare in time use data collections- enables to define weekly measures of time allocated to different categories of activities. The categories we relate to mental load measures are:
\begin{itemize}
    \item\textbf{Household work}: any time spent on meal preparation and clean-up, doing laundry, ironing, dusting, vacuuming, indoor household cleaning, constructing
or repairing objects and goods for one’s own
family, purchasing goods or services for the family, managing family life (planning visits, budgeting, ...). 
\item \textbf{Childcare}: includes activities such as putting to bed/waking
up, helping to eat/drink, helping with bathing/dressing/-
combing/preparing, reading, listening to the child read,
teaching to read, count, write, playing, watching cartoons,
visiting museums, exhibitions, theaters, going to the zoo,
doing artistic activities, watching television, films, series
or programs, browsing the internet, doing manual and
creative activities, going on trips and sports activities, storytelling,
talking, listening, and discussing, organizing activities
or events for the child (e.g., visits, birthday parties,
etc.), helping the child in various activities (e.g., preparing
a backpack, tidying up their things, etc.), supervising
the child without interacting, waiting for the child, accompanying
the child (e.g., to the doctor, etc.), helping
the child with homework, discussing with teachers or
other adults in official roles for both school-related and
extracurricular activities, medicating the child or taking
them to the doctor, outdoor activities.
\item \textbf{Paid work}: it includes all time spent working in the paid sector
in primary, secondary, and overtime jobs, including
time spent working from home, as well as time spent
commuting to/from work. The definition matches that of \citep{gimenez2012trends}.
\end{itemize}
Household work, Childcare and Paid work were measured in terms of average daily hours\footnote{This aggregate is calculated by the formula: (5*weekday+2*weekend)/7}, considering only activities declared as primary activities by the respondent. 
Importantly, thanks to the structure of the TIMES project we observe both partners and can measure - for each of the three categories- the time gap within each couple participating to the survey. In the following the GAP measure is always defined as the difference between the time allocated by the woman and that allocated by the man.
\subsection{Summary statistics}
First, we present summary statistics on socio-demographic characteristics (section 2.3.1), then on mental load variables (section 2.3.2), and lastly on time-use related variables (section 2.3.3).
\subsubsection{Summary statistics on socio-demographic characteristics}
Table \ref{tab:ttest_sum} reports mean difference tests for key demographic characteristics and measures of time use between men and women in the sample. Several significant gender differences emerge. Women are significantly less likely to be employed than men (70.4\% vs. 93.2\%) and less likely to work full time (77.4\% vs. 96.1\%). There are no differences in the likelihood of holding a college degree nor in the years of education.\par
\begin{table}[H]
    \centering
      \caption{Mean difference tests between men and women in socio-demographics.}
      \resizebox{.95\textwidth}{!}{%
    \begin{tabular}{lcccccccc}
        \hline
   &   Women &   Mean &  Men &   Mean &   Diff &   St Err &   P-value \\ 
   \hline
 Employed & 415 & 0.704 & 415 & .981 & -.277 & .024 & .000  \\ 
 Working full time (=1) & 389 & .774 & 279 & .961 & .187 & .024 & .000 \\ 
Education: college (=1) & 415 & 0.393 & 415 & .398 & -.005 & .034 & .888 \\ 
Years of education & 421 & 14.348 & 411 & 14.492 & -.144 & .202 & .477 \\ 
Tot. number of children & 415 & 1.379 & 415 & 1.388 & -.009 & .041 & .812 \\ 
\hline
    \end{tabular}
    }
    \label{tab:ttest_sum}
    \begin{minipage}{\textwidth} 
\scriptsize \textit{Note:} The table reports mean differences between men and women in time use and socio-demographic characteristics. Reported variables include: employment status (1 = employed), full-time work status (1 = full-time), college or higher education (1 = yes), years of education, and the number of children. The variable “Working full-time” is based only on employed individuals (407 men and 292 women), of whom 18 men and 13 women did not report whether they were working full- or part-time.
    \end{minipage} 
\end{table}

\subsubsection{Summary statistics on mental load variables}
In this section, we provide the distribution of our mental load variables and summary statistics of the sample. \par
\begin{figure}[H]
\centering
\caption{Division of organization of household and children's activities.}
{\includegraphics[width=1\linewidth]{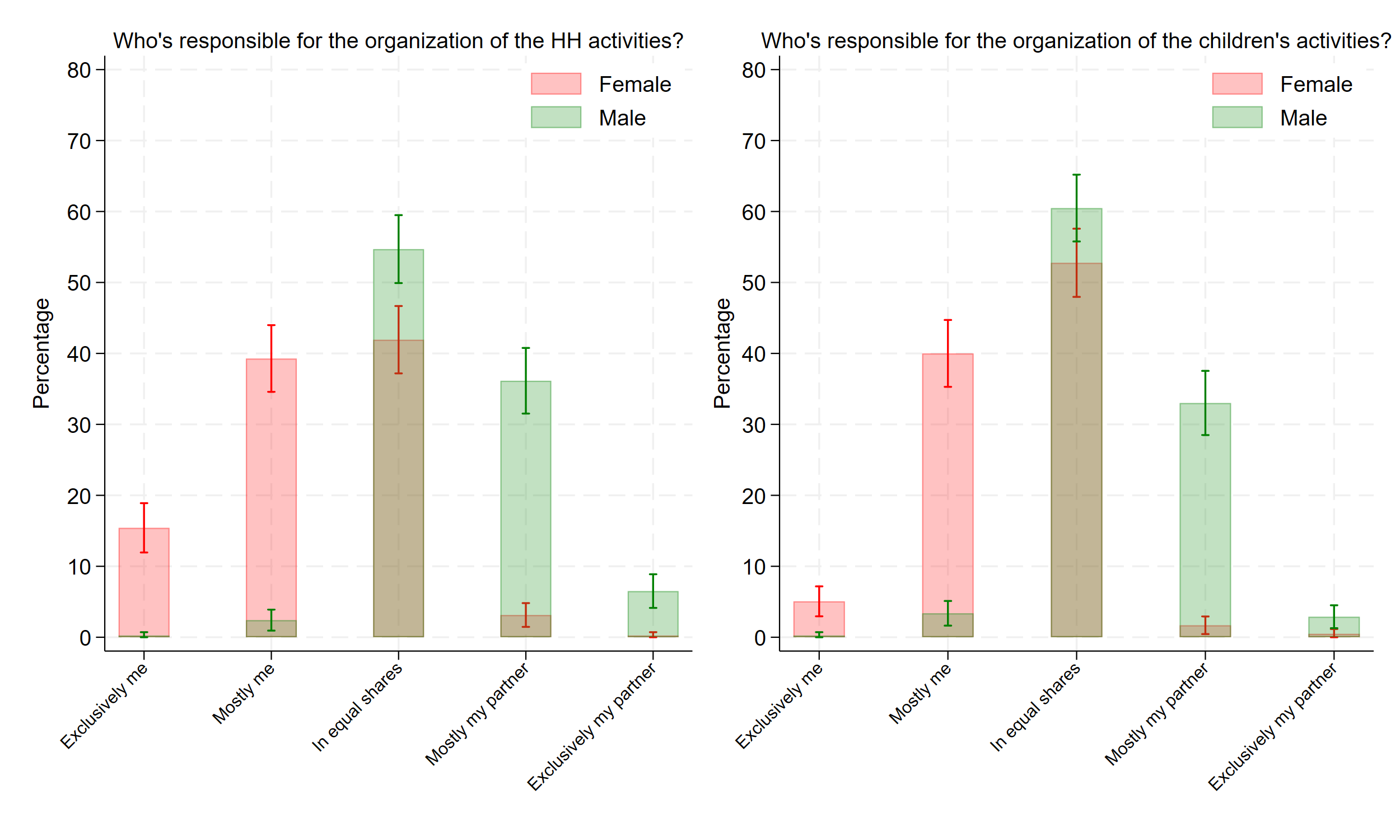}}
\label{fig:CL}
\begin{minipage}{\textwidth} 
\scriptsize The histograms report the distribution of responses from women and men to the questions “\textit{Who is responsible for the organization of household activities?}” and \textit{“Who is responsible for the organization of children’s activities?”} N=830, of which 415 women and 415 men. 
    \end{minipage} 
\end{figure}

 
Figure \ref{fig:CL} reports the distribution of responses to the questions "\textit{Who is responsible for the organization of household activities?}" and "\textit{Who is responsible for the organization of children’s activities?}" separately for women and men (both 415). A clear gendered pattern emerges: women are significantly more likely to report being primarily responsible for these tasks, while men are more likely to report their partner to be in charge. For both household and children’s activities, around 40\% of women report they are “mostly” responsible, with a high share also stating they are “exclusively” responsible. In contrast, the majority of men report that these responsibilities are either shared equally or fall primarily on their partner. The share of respondents selecting “in equal shares” is substantial across genders, but is consistently higher among men, suggesting a perception gap: while many men believe tasks are equally divided, women are more likely to see themselves as the main organizers. These differences are statistically significant, as shown by the non-overlapping confidence intervals in several response categories.

We measure the perceived gender gap in cognitive load in the domain of household and childcare organization, focusing on the \textit{woman as the main organizer} and using two binary indicators. These variables take the value of 1 when women report themselves as the main responsible, and when men report their partner as the main responsible. In our sample, 54.7\% of women state that they are primarily responsible for organizing household tasks, while 42.65\% of men report that their partner holds this role. For the organization of children’s activities, 45\% of women report being the main organizer, compared to 35.9\% of men who attribute this responsibility to their partner. This suggests a possible overestimation by women or underestimation by men, which points to perceptual asymmetries in how organizational responsibilities are recognized within couples.\par 

While our analysis focuses on individual perceptions, having data from both partners enables us to document how these perceptions diverge within the same couple. Table~\ref{tab:CL_partner} in Appendix~A shows the joint distribution of responses to the question \textit{“Who is responsible for the organization of household activities?”} from both women and their male partners. Only about 64\% of couples provide matching answers (i.e., values along the diagonal), while in 36\% of cases, partners disagree. These discrepancies are not symmetric: in 27\% of couples, women report greater responsibility than acknowledged by their partner (upper triangle), while in only 9\% of cases the opposite occurs (lower triangle).
A similar pattern appears for the organization of children’s activities, as shown in Table~\ref{tab:EL_partner} in Appendix~A. Here, close to 70\% of couples provide consistent responses, but when disagreement arises, women again tend to perceive themselves as more responsible than their partner does. This indicates a systematic under-recognition of women's cognitive labor in the family domain.
These findings highlight the limitations of relying on a single respondent. The distribution of cognitive load is not only unequal, but also differently perceived by each partner—revealing mental load as both a gendered and often invisible dimension of domestic life.

\begin{figure}[H]
\centering
\caption{Satisfaction with How the Organization of Household and Childcare Activities Is Shared.}
{\includegraphics[width=.95\linewidth]{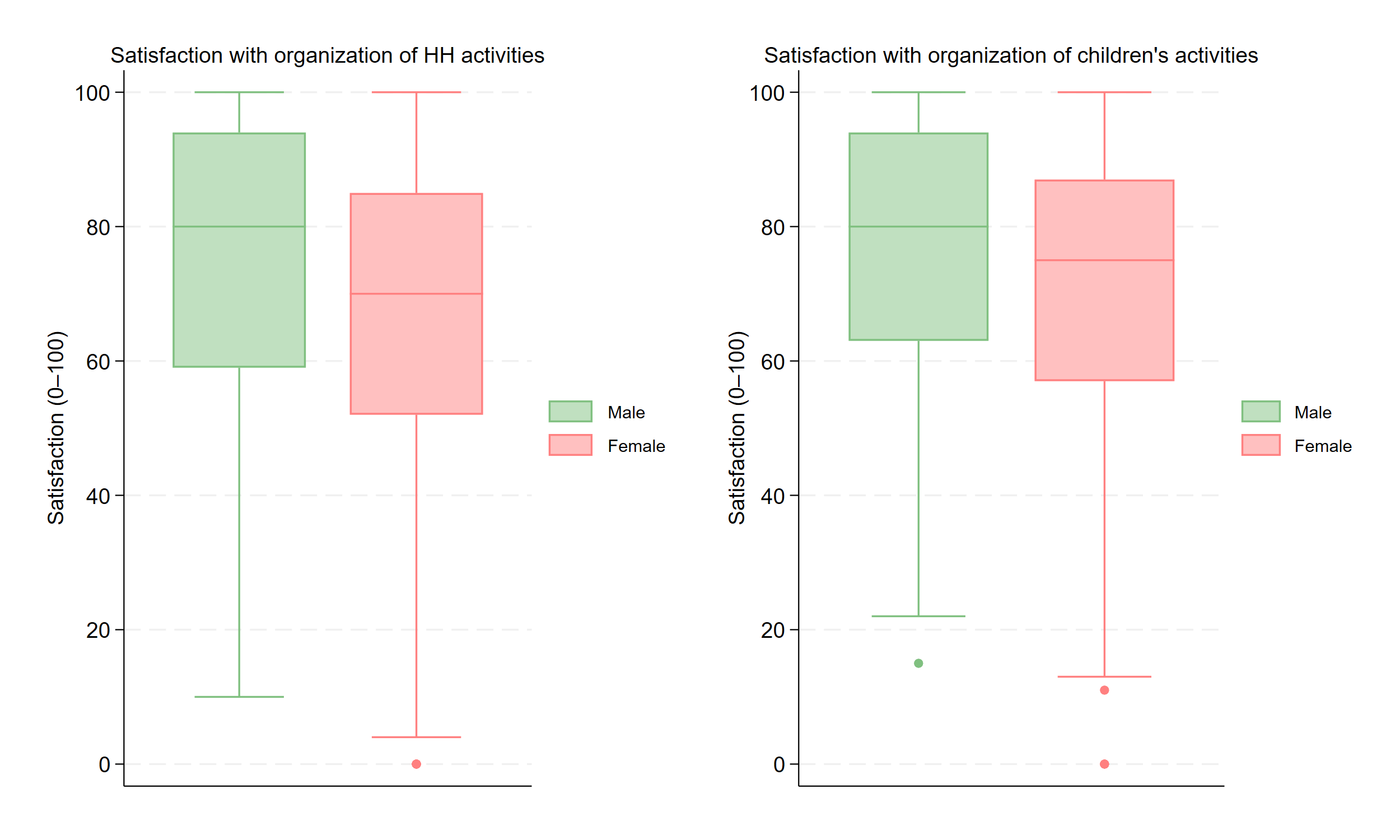}}
\label{fig:satisf}
\begin{minipage}{\textwidth} 
\scriptsize The boxplots display the distribution of satisfaction with the division of organizational responsibilities for household (on the left) and children’s activities (on the right), separately for men and women. Satisfaction is measured on a scale from 0 to 100. N=830, of which 415 women and 415 men.
    \end{minipage} 
\end{figure}
Figure \ref{fig:satisf} presents the distribution of satisfaction levels with the division of organizational responsibilities for household and children’s activities. On average, women report lower satisfaction than men. This pattern is consistent with the earlier finding that women are more likely to identify themselves as the main person responsible for these tasks. The distribution of women’s responses is more dispersed and skewed toward lower satisfaction levels, indicating greater heterogeneity in their experiences. In contrast, men’s satisfaction is more concentrated toward the higher end of the scale, especially regarding the organization of household activities.\par
Figures \ref{fig:CL} and \ref{fig:satisf} reveal an important pattern: although women disproportionately report being the main organizers of both household and children’s activities, this cognitive responsibility does not translate into higher satisfaction. On the contrary, women report significantly lower satisfaction with the division of organizational tasks compared to men (t-test, p-value=0.000). This suggests that the gendered distribution of cognitive labor is not only unequal but also potentially burdensome, particularly for women. The findings highlight a disconnect between responsibility and contentment, suggesting a connection with the emotional costs associated with uneven cognitive load.
\begin{figure}[H]
\centering
\caption{How often do you think about the organization of household and children's activities while at work.}
{\includegraphics[width=.95\linewidth]{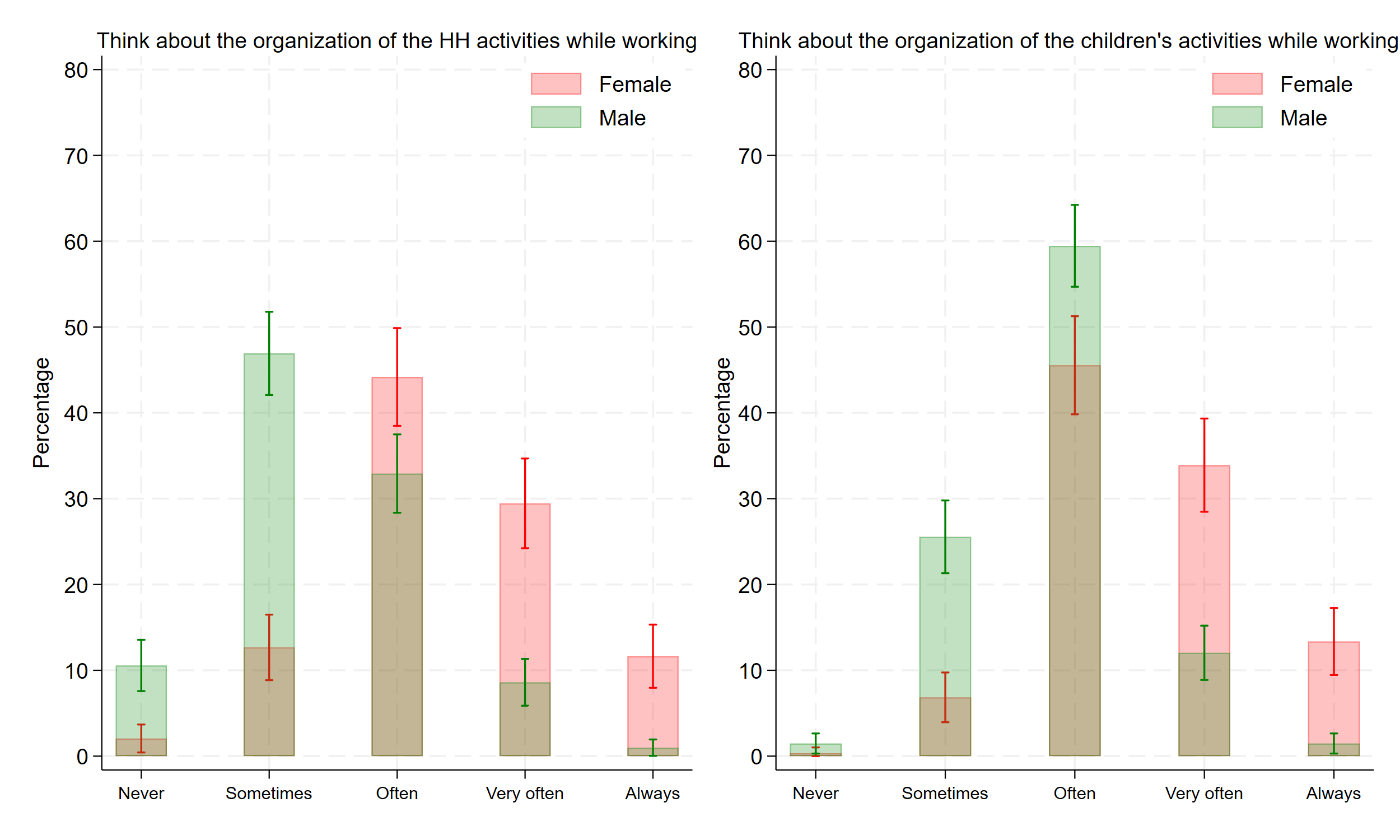}}
\label{fig:think}
\begin{minipage}{\textwidth} 
\scriptsize The bar charts show the distribution of responses to the questions \textit{“How often do you think about the organization of household activities while working?”} and \textit{“How often do you think about the organization of children’s activities while working?”} among employed women and men. Responses were recorded on a five-point Likert scale ranging from Never to Always. N=699, of which 292 women and 407 men. 
    \end{minipage} 
\end{figure}

As an additional dimension of cognitive load, we construct a binary indicator for mental load spillover, coded as 1 if respondents report thinking “very often” or “always” about the organization of household or childcare activities during their workday. Figure \ref{fig:think} reports gender differences in the mental load associated with household and childcare responsibilities during paid work hours. This measure is particularly informative, as it captures the boundaryless nature of mental load: individuals may not be able to fully disconnect from unpaid responsibilities, potentially affecting their productivity or influencing economic decisions \citep{vitellozzi2025invisible}.\par 
Figure \ref{fig:think} reveals striking gender disparities: women are significantly more likely than men to report thinking about both household and childcare organization often, very often, or always while working. These differences are especially pronounced in the context of children's activities, where over 30\% of women report thinking about them very often, compared to just 10\% of men. More precisely, in our sample 41.10\% of employed women report frequently thinking about the organization of household activities while at work, compared to only 9.58\% of men. For childcare activities, 47.26\% of women report doing so, versus 13.51\% of men. These gender gaps are statistically significant (p-value 0.000). 
\begin{table}[H]
    \centering
      \caption{Mean difference tests in feeling of fatigue and responsibility between men and women.}
      \resizebox{.95\textwidth}{!}{%
    \begin{tabular}{lcccccccc}
        \hline
   &   Women &   Mean &  Men &   Mean &   Diff &   St Err &   P-value \\ 
   \hline
 Resp. children's WB  & 415 & 86.048 & 415 & 83.012 & 3.036 & 1.118 & .010 \\  
 Resp. partner's WB  & 415 & 74.393 & 415 & 78.805 & -4.412 & 1.369 & .002 \\ 
 Resp. execution & 415 & 79.788 & 415 & 72.814 & 6.974 & 1.327 & .000 \\
 Resp. family's WB& 415 & 77.532 & 415 & 79.538 & -2.005 & 1.926 & .017 \\ 
 \hline
Responsibility (mean) & 415 & 79.441 & 415 & 78.542 & .898 & 1.049 & .392 \\ 
\hline
&&&&&&&\\
Fatigue children's WB & 415 & 57.998 & 415 & 50.938 & 7.06 & 2.342 & .003 \\ 
 Fatigue partner's WB & 415 & 50.359 & 415 & 49.188 & 1.171 & 2.138 & .584 \\ 
 Fatigue execution  & 415 & 64.082 & 415 & 56.75 & 7.333 & 1.877 & .000 \\ 
 Fatigue family's WB & 415 & 58.983 & 415 & 56.034 & 2.95 & 2.173 & .175 \\ 
 \hline
Fatigue (mean)& 415 & 57.855 & 415 & 53.227 & 4.628 & 1.926 & .017 \\ 
\hline

    \end{tabular}
    }
    \label{tab:ttest_fatigue}
    \begin{minipage}{\textwidth} 
\scriptsize \textit{Note:}The table reports mean differences between men and women in perceived responsibility and fatigue. Each outcome is measured on a scale from 0 to 100 with respect to four domains: children's well-being, partner's well-being, execution of daily activities, and family's overall well-being. The table also includes the average score across these four items.
    \end{minipage} 
\end{table}
Table \ref{tab:ttest_fatigue} presents average scores for men and women on a set of items measuring perceived responsibility and associated fatigue across different domains of family life. Each item is scored on a 0–100 scale, and the table reports the mean values by gender, along with mean differences and results from t-tests for equality of means. On the four items capturing perceived responsibility—covering children’s well-being, partner’s well-being, execution of daily activities, and overall family well-being—women report significantly higher scores than men for children's well-being and for the execution for daily activities. By contrast, men report a higher score of responsibility for their partner well-being of the overall family's well-being, although not significantly different from women.\par
Larger gender differences emerge on the fatigue-related items. Across all four dimensions, women report higher levels of fatigue than men, with statistically significant differences for concerns about children’s well-being and the execution of daily activities. The overall fatigue index is also significantly higher among women. These findings suggest that emotional load is not only unequally distributed but also more intensely experienced by women, particularly in domains closely tied to caregiving and daily management.

\subsubsection{Time use and socio-demographic variables}

\begin{table}[H]
    \centering
      \caption{Mean difference tests between men and women in time use}
      \resizebox{.95\textwidth}{!}{%
    \begin{tabular}{lcccccccc}
        \hline
   &   Women &   Mean &  Men &   Mean &   Diff &   St Err &   P-value \\ 
   \hline
Daily hours spent in HH work & 415 & 2.549 & 415 & .808 & 1.741 & .101 & .000  \\ 
Daily hours spent in childcare & 415 & 2.488 & 415 & 1.59 & .898 & .112 & .000 \\ 
Daily hours spent in paid work & 415 & 3.163 & 415 & 5.619 & -2.455 & .219 & .000 \\ 
\hline
    \end{tabular}
    }
    \label{tab:ttest_time}
    \begin{minipage}{\textwidth} 
\scriptsize \textit{Note:}The table reports mean differences between men and women in time use. Reported variables include the average daily hours spent on housework, childcare, and paid work measured as reported in Section 2.3.
    \end{minipage} 
\end{table}

Table \ref{tab:ttest_time} reports mean difference tests for measures of time use between men and women in the sample. Time use variables show stark gender differences, in line with the pattern documented in the literature based on Italian time use data \citep{barigozzi2023gender}. Women spend significantly more time than men on unpaid domestic work. On average, women spend over 3 hours per day on household tasks compared to less than 1 hour for men, and nearly 4 hours per day on childcare compared to less than 2 hours for men. Conversely, men report spending over 6 hours per day in paid work, while women report only around 2.2 hours. These differences are statistically significant and underscore the unequal division of labor within the household. They also align with gendered patterns of responsibility and cognitive load, which are explored throughout the paper.

\section{Mental load and time spent in household and childcare activities}
To better understand how our mental load measures relate to each other and to time allocation, we estimate bivariate associations between each dimension of mental load and examine their associations both internally and with the considered time use indicators.

\subsection{Cognitive Load}\label{se:cognitive}

\subsubsection*{Division of organization of household and children's activities}
In this section, we relate the probability that the respondent perceives the woman as the main organizer of household and children activities to the time she/he allocates to the respective time use category, as well as to the time use gap between partners.
To this purpose, we estimate the following linear probability models: 
\begingroup
\small
\begin{equation}
\begin{aligned}
    \text{Woman main organizer HH activities}_i = & \ \beta_0 + \beta_1 \text{GAP HH work}_i + \\ + & \beta_2 \text{Time spent HH work}_i + \epsilon_i
\end{aligned}
\label{eq:hh_organizer_model}
\end{equation}
\begin{equation}
\begin{aligned}
    \text{Woman main organizer children's activities}_i = & \ \beta_0 + \beta_1 \text{GAP childcare}_i + \\ + & \beta_2 \text{Time spent childcare}_i + \epsilon_i
\end{aligned}
\label{eq:child_organizer_model}
\end{equation}
\endgroup
\\
\noindent where $\textit{Woman main organizer HH/Children's activities}_i$  is a dummy variable defined as follows: for women, it equals 1 if the respondent reported being the main person responsible for organizing children's or household activities (specifically for women answering "Exclusively me" or "Mostly me"); for men, it equals 1 if the respondent reported their partner as being the main person responsible for these activities (specifically for men answering  "Mostly my partner" or "Exclusively my partner").\footnote{More specifically, we draw on the following questions from the socio-economic questionnaire: \say{\textit{Who is responsible for the organization of household/children's activities?} The possible answers are: \textit{"Exclusively me," "Mostly me," "In equal share," "Mostly my partner,"} and \textit{"Exclusively my partner".}}} The childcare and household work gap is calculated as the difference between the time spent on these categories of ativities by the woman and the time spent by the man measured in hours spent per day. Control variables include the respondent's employment status, educational level, and the number of children. \par 
Each linear probability model is estimated separately for women and men. We estimate three models. In Model 1, we examine the association between the organization of children's activities and household activities with the amount of time devoted to the same activities and gaps in time allocation between partners. In Model 2, we include both categories of activities as covariate in the regression. Lastly, in Model 3, we include the control variables. \par 
\begin{figure}[H]
\centering
\caption{Organization of household and children's activities and time allocation}
{\includegraphics[width=.95\linewidth]{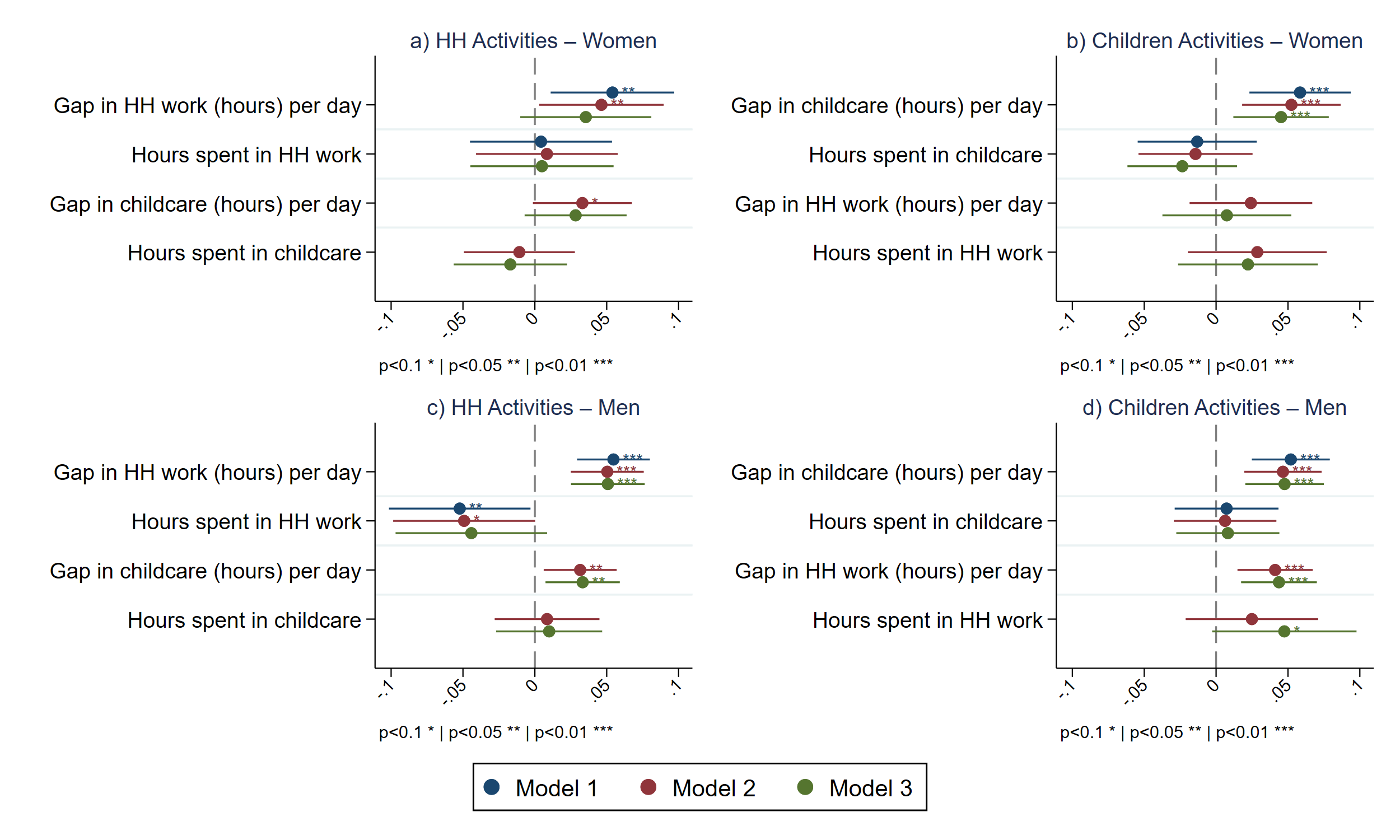}}
\label{fig:CL_time}
\begin{minipage}{\textwidth} 
        \scriptsize \textit{Notes:} *** p$<$0.01; ** p$<$0.05; * p$<$0.10. The reported dependent variable is a binary indicator defined as follows: for women, it equals 1 if the respondent reports being the main person responsible for organizing household and children's activities (i.e., selects “Exclusively me” or “Mostly me”); for men, it equals 1 if the respondent reports that his partner holds the main responsibility for these activities (i.e., selects “Mostly my partner” or “Exclusively my partner”). In the sample, 54.7\% of women report being the main organizer for household activities and 45\% for children's activities, while 42.6\% of men report their partner as the main organizer for household activities and 35.9\% of children's activities. The gap in household (HH) work and childcare is the difference in daily hours spent by women and men (mean HH work gap = 1.74 hours; childcare gap = 2.05 hours). Women spend, on average, 2.54 hours/day on HH work and 2.48 hours/day on childcare, while men spend 0.80 hours/day on HH work and 1.58 hours/day on childcare.
    \end{minipage} 
\end{figure}
Full results from our estimations are reported in Table \ref{tab:CL} of Appendix A.  Figure \ref{fig:CL_time} reports the plots of the coefficients of the models and it shows that the higher the time-use gap between woman and men, the higher is the probability that the woman is identified as the main organizer. This pattern is maintained in both household and childcare activities and is observed in the responses of both women and men. Importantly, this association remains robust across alternative model specifications, including the sequential addition of control variables such as total number of children, employment status, and educational level (Table \ref{tab:CL} in Appendix A).\par
Among women, the larger the gap in time spent on a task compared to their partner, the more likely they are to report themselves as the main organizer of both activities. Among men, the larger the gap in favor of their partner’s time contribution, the more likely they are to report that their partner is the main organizer. In both cases, perceptions of who holds cognitive responsibility within the couple appear to align closely with the observed differences in time allocation. The association is particularly pronounced for childcare, where organizational responsibility is more frequently attributed to the woman as the time gap widens. These results indicate that the management and execution of unpaid household work often coincide. Moreover, they suggest that cognitive responsibility is socially recognized within the couple, thereby reinforcing gendered patterns of time use.

\subsubsection*{Satisfaction with the organization of household and children's activities}
We then focus on partners' satisfaction with the division of the organization of household and children's activities. In the first specification, we include a dummy for whether the woman is the main organizer of these activities to capture the perceived responsibility for planning and coordination, which may affect satisfaction independently of actual time use. In subsequent models, we include time spent in these activities and the gap in time allocation between partners to disentangle the effect of organizational responsibility from that of taking on the practical burden of the tasks. This allows us to separately assess how being responsible for organizing versus carrying out the tasks influences satisfaction with their division. We estimate the following specifications: \par 
\begin{equation}
\begin{aligned}
\text{Satisf.\ organization\_CHILD}_i 
 &= \beta_0
    + \beta_1\,\text{Woman\_main\_organizer\_CHILD}_i \\
 &\quad + \beta_2\,\text{Time spent childcare}_i 
    + \beta_3\,\text{GAP childcare}_i + \epsilon_i
\end{aligned}
\label{eq:satisf_CHILD_model}
\end{equation}

\begin{equation}
\begin{aligned}
\text{Satisf.\ organization\_HH}_i 
 &= \beta_0
 + \beta_1\,\text{Woman\_main\_organizer\_HH}_i \\
 &\quad + \beta_2\,\text{Time spent HH work}_i
 + \beta_3\,\text{GAP HH work}_i + \epsilon_i
\end{aligned}
\label{eq:satisf_HH_model}
\end{equation}
\\
\noindent Each model is estimated separately for women and men. The dependent variable ranges from 0 to 100, with higher values indicating greater satisfaction.\footnote{Specifically, we refer to the question in the socio-economic questionnaire: ``\textit{On a scale from 0 to 100, how satisfied are you with the division of the organization of children's/household activities?}''
} \par 
We estimate four models. In Model 1, we include a binary variable indicating whether the woman is the main person responsible for organizing the activity. This captures the potential cognitive or managerial burden associated with being in charge of planning and coordination. In Model 2, we add the number of hours spent on the activity to isolate the effect of time allocation from that of organizational responsibility. This allows us to disentangle whether satisfaction is more closely linked to doing the task or managing it. In Model 3, we include the within-couple gap in time allocation, which reflects perceived fairness or imbalance in contributions. Finally, Model 4 adds socio-demographic controls (employment status, education, number of children) to account for observable heterogeneity across households.\par

\begin{figure}[H]
\centering
\caption{Satisfaction with organization of household and children's activities}
{\includegraphics[width=.95\linewidth]{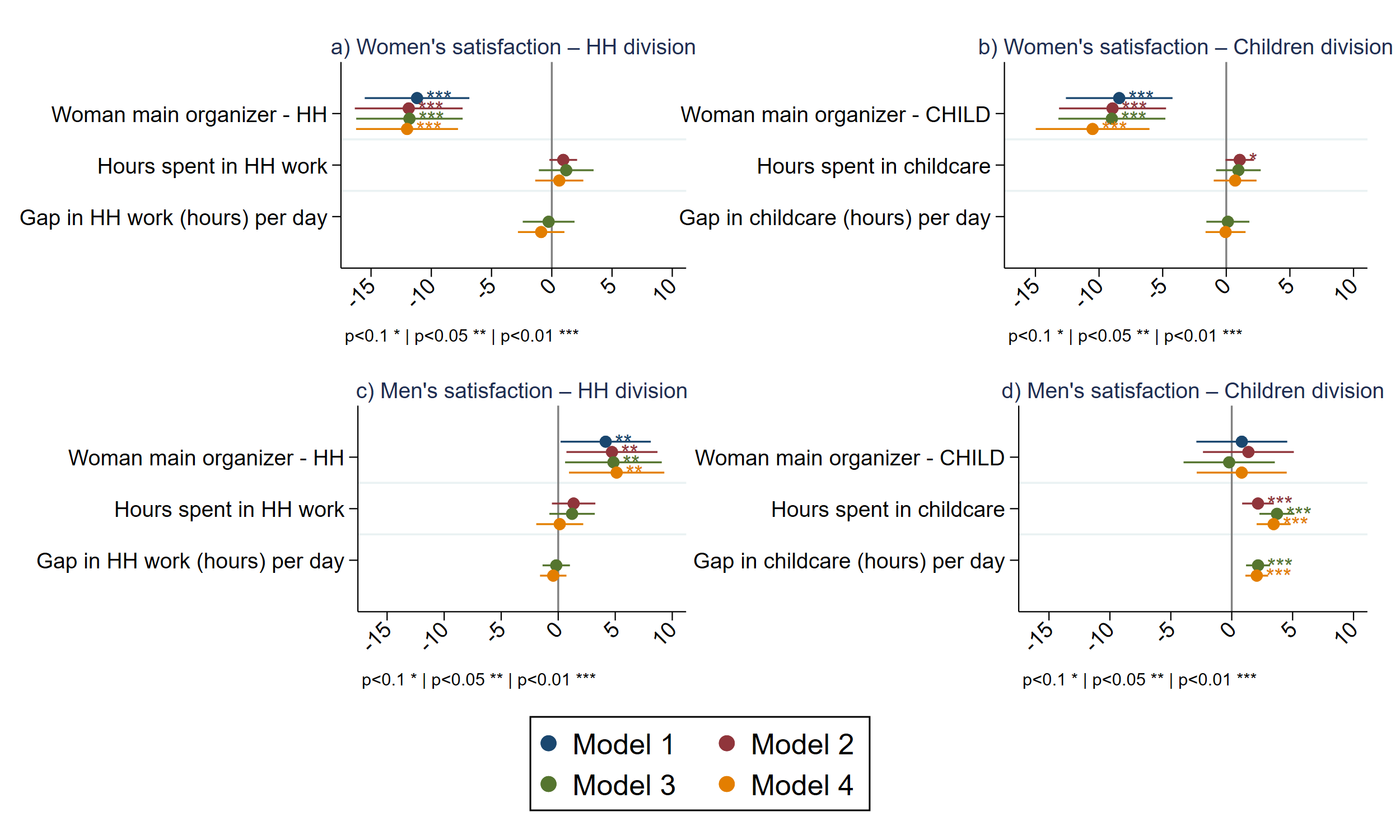}}
\label{fig:satisf_time}
\begin{minipage}{\textwidth} 
        \scriptsize \textit{Notes:} *** p$<$0.01; ** p$<$0.05; * p$<$0.10. *** p$<$0.01; ** p$<$0.05; * p$<$0.10. The dependent variables are satisfaction scales (0–100) for the questions: "How satisfied are you with the division of the organization of HH activities?" (women mean = 67.27; men mean = 75.55) and "How satisfied are you with the division of the organization of children's activities?" (women mean = 71.48; men mean = 77.40). The gap in household (HH) work and childcare is the difference in daily hours spent by women and men (mean HH work gap = 1.74 hours; childcare gap = 2.05 hours). Women spend, on average, 2.54 hours/day on HH work and 2.48 hours/day on childcare, while men spend 0.80 hours/day on HH work and 1.58 hours/day on childcare. The reported estimates are for the four models in which we gradually include additional control variables. 
    \end{minipage} 

\end{figure}

Full results from our estimations are reported in Table \ref{tab:satisf} of Appendix A.
Figure \ref{fig:satisf_time} reports the plots of the coefficients of the models. Among women, there is a negative association between being the main responsible for organizing household or children's activities and reported satisfaction. This association is particularly evident as the time-use gap increases: when women spend substantially more time than their partner on unpaid domestic or childcare tasks, satisfaction with the division of responsibilities tends to be lower. In other words, both being the main organizer and spending more time on unpaid work are associated with lower satisfaction for women.\par
For men, satisfaction with the organization of household activities tends to be positively associated with their female partner being the main organizer. By contrast, their satisfaction with the organization of children's activities is positively related to the time they spent in childcare. \par
These patterns highlight the gendered asymmetries in how organizational responsibility and time dedicated to household chores and childcare shape satisfaction. 
These associations remain robust across alternative model specifications, including the sequential addition of control variables such as total number of children, employment status, and educational level (Table \ref{tab:satisf} in Appendix A). \par

\subsubsection*{Thinking about the organization of household and children's activities}
Lastly, we investigate the third dimension of cognitive load: its spillover into the workplace. Specifically, we study how often individuals report thinking “very often” or “always” about the organization of household and childcare activities while at work. As formalized in Equations \ref{eq:hh_organizer_model_1} and \ref{eq:child_organizer_model_1}, we estimate four nested models to examine the correlates of this mental preoccupation.
\begin{equation}
\begin{aligned}
    \text{Think about HH activities}_i = & \ \beta_0 + \beta_1 \text{Woman main organizer HH}_i + \epsilon_i
\end{aligned}
\label{eq:hh_organizer_model_1}
\end{equation}
\begin{equation}
\begin{aligned}
\text{Think about childcare}_i =\ & \beta_0 + \beta_1\, \text{Woman main organizer childcare}_i + \epsilon_i
\end{aligned}
\label{eq:child_organizer_model_1}
\end{equation} 
\\
\noindent We begin by analyzing the association between thinking about domestic organization during work hours and being the main person responsible for such organization. This captures the boundaryless nature of mental load—the extent to which organizational responsibilities spill beyond the home and into paid work contexts. To this end, we only include in this analysis respondents who report to have a job. Specifically, the sample includes 70.3\% (N=292/415) working women and 98.07\% (N=407/415) working men. \par 
In Model 1, we include only the indicator for being (or, for men, having a partner who is) the main organizer of household or childcare tasks. In Model 2, we add time spent on the relevant unpaid task and in paid work. This allows us to assess whether the correlation between organizational responsibility and spillover remains after accounting for actual time investments. In Model 3, we add the within-couple time gap in unpaid work, which may reflect imbalances in task-sharing and influence stress or perceived fairness. Finally, Model 4 includes standard socio-demographic controls (number of children, education, employment status). \par
\begin{figure}[H]
\centering
\caption{Thinking about the organization of children's and HH activities at work}
{\includegraphics[width=0.9\linewidth]{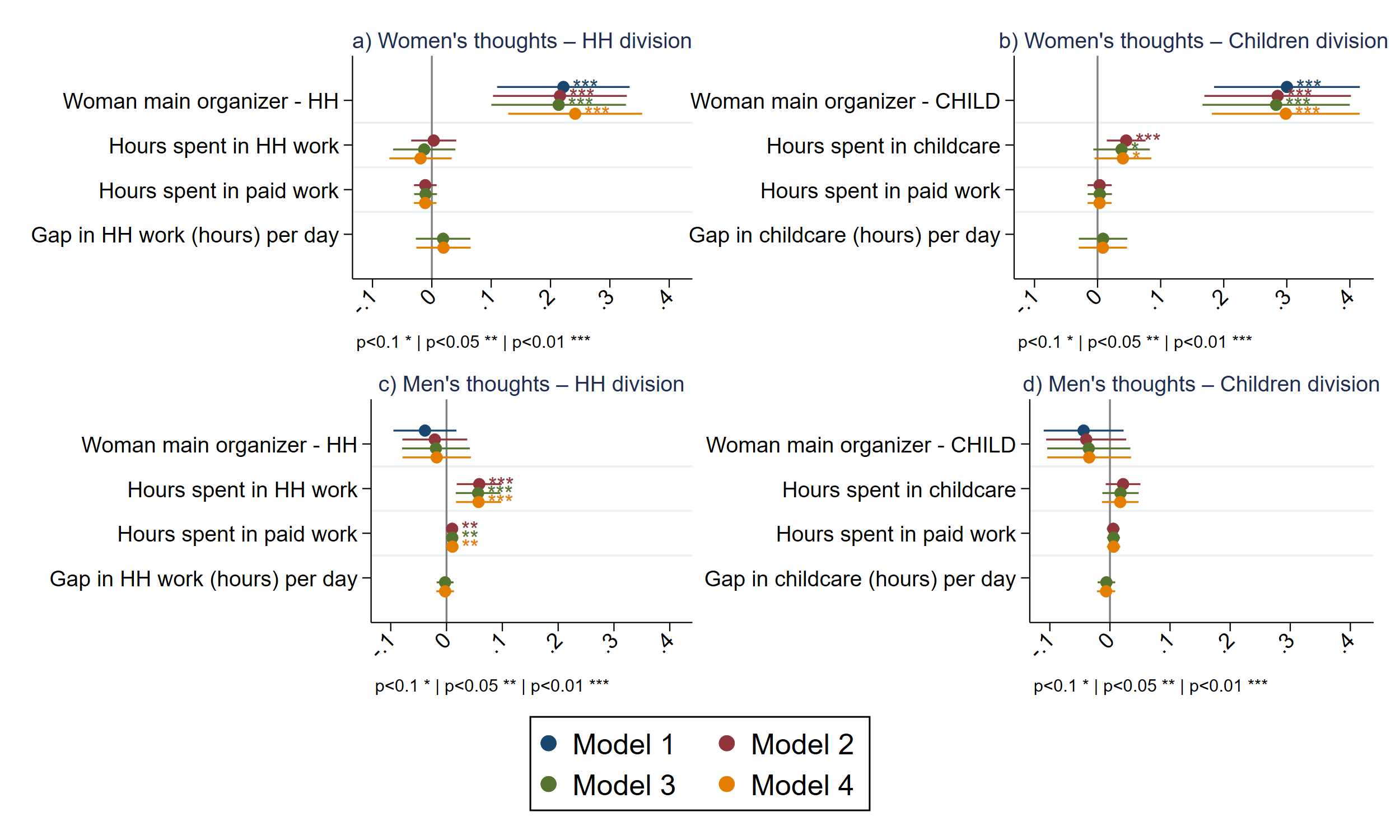}}
\label{fig:think_time}
\begin{minipage}{\textwidth} 
        \scriptsize \textit{Notes:} *** p$<$0.01; ** p$<$0.05; * p$<$0.10. The models are estimated on respondents who self-report to have a job (70.3\% (N=292/415) working women and 98.07\% (N=408/415) working men.) 
        The dependent variable is a binary indicator equal to 1 if the respondent reports thinking “very often” or “always” about the organization of household or children’s activities while at work. In the sample, 41.1\% of women report frequent thoughts about household organization at work, and 47\% do so for childcare. Among men, the corresponding shares are 9.58\% and 13.5\%, respectively. The gap in household (HH) work and childcare is the difference in daily hours spent by women and men (mean HH work gap = 1.74 hours; childcare gap = 2.05 hours). Women spend, on average, 2.54 hours/day on HH work and 2.48 hours/day on childcare, while men spend 0.80 hours/day on HH work and 1.58 hours/day on childcare.
    \end{minipage} 

\end{figure}

Full results from our estimations are reported in Table \ref{tab:think} of Appendix A.
As shown in Figure \ref{fig:think_time}, women who report being the main organizer of household activities (Panel a) or childcare (Panel b) are also more likely to report thinking “very often” or “always” about these responsibilities during their workday. Among men, those who report that their partner is the main organizer (Panel c and d) are, on average, less likely to report thinking frequently about organizational responsibilities at work, although this relationship is not statistically significant. Conversely, the more time men spend on household work, the more likely they are to report thinking about household organization while working (Panel c), suggesting a potential experiential component to the mental load for men. This association remains robust across alternative model specifications, including the sequential addition of control variables such as total number of children, employment status, and educational level (Table \ref{tab:think} in Appendix A). \par

These patterns highlights the risk that mental load generates negative spillovers on productivity and engagement at work—particularly for women—by taxing cognitive resources during paid work hours. Organizational responsibility does not end with initial planning; it requires sustained mental effort to monitor whether tasks are being executed properly, anticipate potential disruptions, and adjust plans as needed. This ongoing, invisible process of tracking and re-planning can fragment attention, increase stress, diminish availability for demanding work-related tasks, and reduce productivity \citep{offer2014costs, dean2022mental, vitellozzi2025invisible}. \par

Notably, the effort involved in organizing family tasks and the time spent carrying them out are complementary. Our analysis shows that women who report being the primary organizers also devote more time than their partners to household and childcare tasks. Moving to double-earner couples, the cognitive burden operates alongside the well-documented “second shift” \citep{hochschild2012second}, wherein working women disproportionately shoulder the demands of domestic labor after formal work hours. Empirical evidence consistently confirms that women spend more time on unpaid domestic tasks than men, even in dual-earner households (\citealt{bittman2003does};\citealt{gimenez2012trends}; \citealt{barigozzi2023gender};\citealt{ferrant2014unpaid}). Mental load intensifies this inequality by extending the burden into cognitive space, creating a double strain of time and attention. 


\subsection{Emotional Load}
In this section, we explore the associationbetween perceived emotional load, satisfaction with the division of the organization of household and childcare responsibilities, and the time devoted to these activities.\par
To measure emotional load, we rely on four items from the socio-economic questionnaire in which respondents report how often they feel exhausted when thinking about: (i) their children’s well-being, (ii) their partner’s well-being, (iii) their family’s overall well-being, and (iv) the execution of daily tasks. We take the average of these items to construct a summary measure of emotional fatigue, which serves as our dependent variable.\par
We are particularly interested in whether emotional fatigue is correlated with dissatisfaction over how organizational responsibilities are divided within the household. Satisfaction is interpreted here as a proxy for whether the organizational burden is perceived as fair or sustainable—thus potentially indicating whether cognitive responsibilities translate into emotional strain.\par
We estimate the following regression:  \par

\begin{equation}
\begin{aligned}
    \text{Fatigue}_i &= \beta_0 + \beta_1 \text{Satisf\_HH}_i + \\
    &\quad + \beta_2 \text{Satisf\_CHILD}_i + \epsilon_i
\end{aligned}
\label{eq:fatigue_model}
\end{equation} 
\\
In Model 1, we examine the correlation between satisfaction with the division of organizational responsibilities and emotional fatigue. In Model 2, we add time allocation variables—hours spent on household and childcare tasks, as well as time in paid work. This allows us to distinguish between emotional strain that stems from perceived organizational responsibility (cognitive load) and that which may be driven by the physical time demands of task execution. Model 3  includes standard socio-demographic controls.
\par 
\begin{figure}[H]
\centering
\caption{Feeling of fatigue and satisfaction with organization of children's and HH activities}
{\includegraphics[width=1\linewidth]{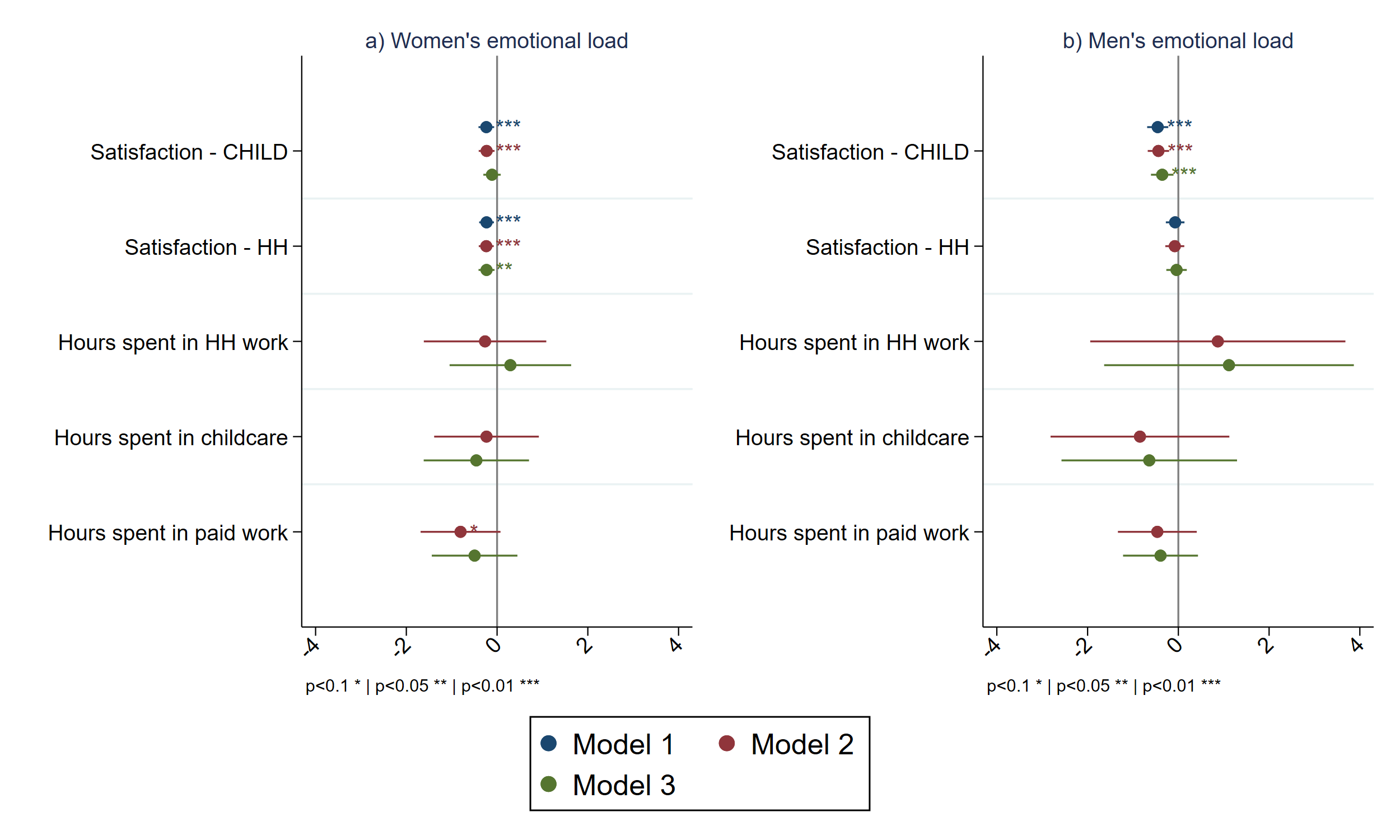}}
\label{fig:EL_time}
\begin{minipage}{\textwidth} 
        \scriptsize \textit{Notes:} *** p$<$0.01; ** p$<$0.05; * p$<$0.10. The feeling of fatigue is measured as the mean of four variables measuring, on a scale from 0 to 100, the level of fatigue for four dimensions: children's well-being; partner's well-being; execution of daily activities; and overall family's well-being (women mean = 60.3; men mean = 57). Control variables include a satisfaction scale ranging from 0 to 100 of the division of the organization of household and children's activities; the time spent on average in housework and childcare (women spend, on average, 2.84 hours/day on HH work and 3.06 hours/day on childcare, while men spend 0.78 hours/day on HH work and 1.64 hours/day on childcare); the average time spent in paid work; the employment status of the respondent; the total number of children; whether the respondent has a college degree or not; and the region of residence. The reported estimates are for the four models in which we gradually include additional control variables. 
    \end{minipage} 

\end{figure}
Full results from our estimations are reported in Table \ref{tab:fatigue} of Appendix A. Figure \ref{fig:EL_time} shows the associations between perceived emotional fatigue and a set of key predictors, separately for women (Panel a) and men (Panel b). The feeling of fatigue is measured as the average of four items related to concern for children, partner, family, and daily tasks. The models progressively add controls for time use and socio-demographics.\par
For women, emotional fatigue is strongly and negatively associated with satisfaction regarding the organization of both household and children’s activities. This relationship is robust across all three models and suggests that women who are less satisfied with how these tasks are organized report significantly higher levels of fatigue. Time use variables also play a role. The amount of time women spend in childcare is positively associated with emotional fatigue, even after controlling for satisfaction. This indicates that the actual burden of execution (the physical component of unpaid labor) contributes to emotional strain beyond cognitive or organizational responsibilities. Time spent in paid work also correlates positively with emotional fatigue, highlighting the potential cumulative toll of balancing professional and unpaid domestic responsibilities.\par
For men, the patterns are generally weaker and less statistically significant. While satisfaction with the division of household and childcare organization is negatively associated with emotional fatigue in Model 1, these associations attenuate as time use and controls are added. Moreover, none of the time use variables show strong or consistent correlations with men’s emotional fatigue. This suggests that men may be less affected—emotionally and cognitively—by the way unpaid responsibilities are organized and executed within the household, at least as captured by this set of measures.\par 
Importantly, this association remains robust across alternative model specifications, including the sequential addition of control variables such as the number of children, the educational level, and the employment status (Table \ref{tab:fatigue} in Appendix A).\par
Taken together with our earlier findings on cognitive labor, these results highlight the importance of jointly measuring both the emotional and cognitive dimensions of mental load. While cognitive load captures the burden of planning, monitoring, and coordinating family activities, emotional load reflects the psychological toll of caring and concern—particularly when individuals feel unsupported or dissatisfied with how managing tasks are shared. The gendered pattern we observe—where both the mental planning and emotional strain are disproportionately borne by women—reinforces the notion that mental load is multifaceted and cumulative. Capturing these distinct but interrelated aspects is essential for a fuller understanding of how unpaid labor contributes to gender inequality in both the private and professional spheres.

\subsection{Focus on working women}

Understanding how mental load manifests among working women is particularly important, as they must navigate the dual demands of paid work and unpaid domestic responsibilities. To this end, we replicate all the analyses presented in previous sections, restricting the sample to employed women. Full results are reported in Tables \ref{tab:mean_emp}, \ref{tab:CL_emp}, \ref{tab:satisf_emp}, and \ref{tab:fatigue_emp} in Appendix A. In this section, we summarize the most relevant trends.

When comparing the key indicators between employed and non-employed women (see Table \ref{tab:mean_emp} in Appendix A), clear and substantial differences emerge in the distribution of organizational responsibility. Non-employed women are far more likely to identify themselves as the main organizer of both household and childcare tasks: 74\% versus 47\% for household work, and 71\% versus 34\% for childcare. These differences are large and statistically significant at the 1\% level, pointing to a meaningful reallocation of cognitive duties when women engage in paid employment.

However, this shift in responsibility is not matched by a clear increase in satisfaction with how planning tasks are shared. Employed women report slightly lower satisfaction levels—between 2.5 and 4.3 points lower on a 0–100 scale—but these differences are only marginally significant. This suggests that a reduced role in organizing tasks does not necessarily lead to a stronger perception of fairness or balance within the household.

Emotional fatigue remains largely stable across employment status. Although non-employed women report slightly higher levels of emotional strain, the difference (2.4 points) is small and not statistically significant. This indicates that, unlike cognitive load, emotional burdens appear less responsive to shifts in time use or task delegation.

\subsubsection*{Cognitive load}

We now turn to the relationship between time-use gaps and perceived cognitive responsibility, replicating the analysis from Section \ref{se:cognitive} while restricting the sample to women and accounting for employment status. Full results are presented in Tables~\ref{tab:CL_emp} and~\ref{tab:satisf_emp}, and illustrated in Figures~\ref{fig:cl_emp} and~\ref{fig:satisf_emp} in Appendix~A.

Across both employed and non-employed women, we observe a positive association between time-use gaps and the probability of being identified as the main organizer of household activities. However, this relationship is stronger among employed women. As the time-use gap increases—i.e., as the woman spends more time than her partner on domestic work—the likelihood that she reports being the main organizer rises more steeply. This sharper alignment may reflect the greater salience or perceived burden of planning when time is scarce and competing demands from paid work are present.

Among non-employed women, the association is weaker, suggesting that time spent on domestic tasks is less strongly tied to perceptions of organizational responsibility in the absence of paid work.

A similar dynamic is observed when considering satisfaction with the division of organizational tasks. In both groups, satisfaction tends to decrease as the time-use gap grows, but this decline is more marked among employed women—particularly for childcare responsibilities. This pattern suggests that time imbalances may be more frustrating for those juggling paid work, while non-employed women may experience or interpret these imbalances with more flexibility or acceptance.

Overall, these findings indicate that employment status conditions how time inputs are translated into both perceived responsibility and satisfaction. The alignment between time spent and perceived cognitive responsibility appears stronger—and its consequences more pronounced—among women who are also engaged in paid work.

\subsubsection*{Emotional load}

We now examine the determinants of emotional fatigue, comparing employed and non-employed women. Full results are presented in Table~\ref{tab:fatigue_emp} and illustrated in Figure~\ref{fig:fatigue_emp_vs_unemp} in Appendix~A.

While average levels of emotional fatigue are similar across the two groups, the factors associated with this burden differ depending on employment status. Among employed women, emotional fatigue is significantly associated with dissatisfaction regarding the organization of household activities, suggesting that poor coordination of domestic responsibilities contributes to emotional strain when combined with paid work. By contrast, dissatisfaction with the organization of children’s activities does not show a significant link to emotional fatigue in this group.

Among non-employed women, the pattern is reversed. Emotional fatigue is more strongly linked to dissatisfaction with childcare planning, while the association with household task coordination is weaker and not statistically significant. This may reflect the centrality of caregiving tasks in the daily routines of women not engaged in paid employment, making disorganization in this domain more emotionally taxing.

 To sum up, although working women tend to share organizational duties more equally with their partners compared to non-employed women, this shift does not translate into greater satisfaction. In fact, employed women appear more sensitive to imbalances in time use and planning responsibilities, particularly in the domain of childcare. This suggests that for women juggling work and family, unequal distribution of organizational tasks is a significant source of dissatisfaction—possibly because they expect not only help in performing tasks, but also greater involvement from their partners in planning and managing them.

\subsection{The source of mental load: time spent or within couple time-gap?}
The distinction between absolute time and the within-couple time gap is conceptually relevant for understanding mental load. Absolute time captures the executional burden—the physical hours spent on unpaid work—which might intuitively link to mental strain. However, the time gap reflects inequality or imbalance, which is central to mental load as a relational and gendered experience. A woman may spend many hours on childcare, but it is the fact that she does more than her partner that signals disproportionate cognitive responsibility and potentially affects well-being and satisfaction. Both dimensions—burden and imbalance—contribute to the lived experience of mental load and thus must be considered jointly in empirical analyses.\par
Based on the regression results of Section 3.1 displayed in Table \ref{tab:CL} in Appendix A, Tables \ref{table4} and \ref{table5} present F-tests assessing the relationship between self-reported organizational responsibility (i.e., being the main organizer of household or children’s activities) and the two key dimensions of time use: (i) the absolute time spent on the activity and (ii) the gender gap in time use between partners.\par
To interpret the outcome of the test, it is useful to rewrite the models estimated in Section 3.1  with a general notation encompassing both types of activities,  but differentiating the models for female and males respondents. For the female sample we have:\par
\begingroup
\small
\begin{equation}
\begin{aligned}
    \text{Woman main organizer}_i = & \ \beta_0 + \beta_1 (\text{woman time}_i-\text{man time}_i )+ \\ + & \beta_2 \text{woman time}_i + \epsilon_i
\end{aligned}
\label{eq:test woman}
\end{equation}
While for the male sample the model becomes:
\begingroup
\small
\begin{equation}
\begin{aligned}
    \text{Woman main organizer}_i = & \ \beta_0 + \beta_1 (\text{woman time}_i-\text{man time}_i )+ \\ + & \beta_2 \text{man time}_i + \epsilon_i
\end{aligned}
\label{eq:test man}
\end{equation} 
\par
Consider Table \ref{table4} first. For women, the gender gap in household work is significantly associated with being the main organizer (Model 1 and 2 - HP2), but the woman’s own absolute time spent is not (HP1). This suggests that women’s perception of organizational responsibility is shaped more by how much more they do relative to their partner than by how much time they spend per se. For female respondents, HP3 corresponds to the assumption that only the man's time matters,  for the mental load of the woman, and this hypothesis is strongly rejected.\par
For men, both the absolute time spent in household work (Model 1 and 2 - HP1) and the time gap significantly predict whether men report their partner as the main organizer. This indicates that men interpret cognitive responsibility through both lenses—how much they do as well as how unequally it is shared. For males respondents HP3 corresponds to the assumption that only woman's time matter, and this is strongly rejected.

\begin{table}[H]
\caption{Hypotheses testing - organization with division of HH activities and time allocation}
\resizebox{\textwidth}{!}{%
\begin{tabular}{lccc}
\hline
& Model 1 & Model 2 & Model 3 \\
\hline
Female sample & Prob $>$ F & Prob $>$ F & Prob $>$ F \\
\hline
HP 1: Time spent in HH work ($\beta_2$) = 0 & 0.8644 &  0.7384 & 0.8452\\
HP 2: Gap in HH work ($\beta_1$) = 0 & 0.0140 & 0.0360 & 0.1271\\
HP 3: Gap in HH work ($\beta_1$) = $-$ Time spent in HH work ($\beta_2$) & 0.0000 & 0.0000 & 0.0040\\
\hline
Male sample & Prob $>$ F & Prob $>$ F & Prob $>$ F \\
\hline
HP 1: Time spent in HH work ($\beta_2$) = 0 &  0.0375 & 0.0509 & 0.1002\\
HP 2: Gap in HH work ($\beta_1$) = 0 & 0.0000 & 0.0001 & 0.0001\\
HP 3: Gap in HH work ($\beta_1$) = Time spent in HH work ($\beta_2$) & 0.0000 & 0.0000 & 0.0001\\
\hline
\end{tabular}
}
\label{table4}
\end{table}

Table \ref{table5} mirrors the analysis for children’s activities and confirms a similar pattern. For women, in all models, the gender gap in childcare is the significant predictor of perceived organizational responsibility, while their own time spent in childcare is not. This pattern supports the idea that relative contributions, not just workload, shape perceptions of mental load. HP3 is not rejected in Model 3, providing some indication that the man time dedicated to childcare can be a driving force diminishing the woman mental load.\par
For men, the greater time in childcare gap strongly correlates with men identifying their partner as the main organizer, whereas the absolute own time does not contribute to the mental load of their female partner. HP3 that only the woman time matters is again rejected.

\begin{table}[H]
\centering
\caption{Hypotheses testing - organization with division of children's activities and time allocation}
\resizebox{\textwidth}{!}{%
\begin{tabular}{lccc}
\hline
& Model 1 & Model 2 & Model 3 \\ \hline
Female sample & Prob $>$ F & Prob $>$ F & Prob $>$ F \\
\hline
HP 1: Time spent in childcare ($\beta_2$) = 0 & 0.5332 & 0.4787 & 0.2257\\
HP 2: Gap in childcare ($\beta_1$) = 0 & 0.0012 & 0.0028 &  0.0077\\
HP 3: Gap in childcare ($\beta_1$) = $-$ Time spent in childcare ($\beta_2$) & 0.0013 & 0.0052 & 0.1134\\
\hline
Male sample & Prob $>$ F & Prob $>$ F & Prob $>$ F \\
\hline
HP 1: Time spent in childcare ($\beta_2$) = 0 & 0.6915 & 0.7297 & 0.6545\\
HP 2: Gap in childcare ($\beta_1$) = 0 & 0.0002 & 0.0007 & 0.0007\\
HP 3: Gap in childcare ($\beta_1$) = Time spent in childcare ($\beta_2$) & 0.0031 & 0.0082 & 0.0096\\
\hline
\end{tabular}
}
\label{table5}
\end{table}


\section{Mental Load: the State of the Art}\label{sec:related.liteature}

While the concept of \textit{mental load} has gained increasing attention across disciplines, its empirical measurement remains inconsistent and fragmented. A growing body of work has begun to define and quantify this form of invisible labor, yet existing approaches vary significantly in terms of definition, operationalization, and scalability.

Early contributions, such as \citet{daminger2019cognitive}, offer a rich qualitative framework that identifies mental load as composed of anticipation, planning, decision-making, and monitoring. However, her method, based on in-depth interviews with heterosexual couples, remains exploratory and non-scalable, lacking a standardized survey tool for broader implementation.
 \citet{wayne2023s} develop a structured and validated 9-item scale to measure what they term \textit{invisible family load}, disaggregated into managerial, cognitive, and emotional dimensions. Their instrument is modular and suitable for survey use, and importantly, it includes emotional labor explicitly. Nonetheless, their data is collected from only one respondent per household, limiting the ability to compare perceptions within couples -a central feature of our approach.

\citet{dean2022mental} offer a conceptual contribution, advocating for a redefinition of mental load that explicitly includes both cognitive and emotional labor. They highlight the invisible, enduring, and boundaryless nature of this form of work and suggest its integration into time-use surveys. However, their work is theoretical and does not propose a directly implementable empirical strategy. In contrast, \citet{offer2014costs} uses an experience sampling method to track mental labor in real time across different life domains. While this provides valuable insights on gendered spillovers between work and family, the method is intensive, open-ended, and difficult to scale for use in large, general-purpose surveys.

\citet{haupt2024gendered} use existing time-use retrospective questions from the Generations and Gender Survey (GGS) as proxies for cognitive household labor. Their approach innovatively links mental load to family--work conflict but does so using indirect attribution, rather than directly capturing individuals' perceptions of mental effort and planning responsibilities.

Finally, \citet{vitellozzi2025invisible} take an experimental approach, priming mental load via videos and time diaries\footnote{Time use data were collected through 24-hour recall diaries administered during the interview, using a module adapted from the Women’s Empowerment in Agriculture Index (WEAI) \citep{alkire2013women}.} to measure its impact on productivity in low-income areas in Kenya. While their study provides valuable insights into the economic consequences of mental load, it is designed for experimental contexts and not intended for generalizable survey use.

Our contribution differs from previous research along several key dimensions.

From a conceptual point of view, we define mental load as the cognitive and emotional preoccupations related to organizing, anticipating, and coordinating household tasks, integrating both instrumental and affective components.

We implement a new set of structured survey items that ask both partners in a couple who is responsible for thinking about and planning recurring domestic activities. This dual-perspective approach allows us to quantify intra-household divergences in mental load attribution. In other words, by combining self- and partner-reported data, our framework enables the study of \textit{perception gaps} in the attribution of invisible labor---a novel angle largely absent in existing quantitative literature.
    
Our instrument is designed to be modular and easily integrated into existing survey infrastructure. In contrast to qualitative interviews or experience sampling, our method is replicable and suitable for large-scale population studies.

In summary, while previous research has deepened our understanding of mental load, this paper introduces a scalable, survey-ready tool that captures both the cognitive and emotional dimensions of mental load from the perspective of both partners in a couple. This dual-respondent structure opens new avenues for studying how mental load is perceived, distributed, and justified within families.

This study represents an initial step toward analyzing the important and often overlooked dimension of gender inequality in mental load. In the present paper, we make only partial use of the rich information collected from both partners. Specifically, among the explanatory variables, we focus on gender gaps in time use measured at the couple level.
The important issue of potential tensions between partners in managing family responsibilities—along with discrepancies in their perceptions of the division of labor and the associated burden—falls outside the main scope of this analysis. An anticipation in this direction is provided in Appendix A; on this point, see also our discussion in the concluding section.

\section{Conclusion}
This paper offers a novel analysis of mental load within families leveraging a unique dataset that includes both partners in 415 heterosexual couples with children under 11. 
Drawing on recent theoretical developments, we distinguish between cognitive and emotional dimensions of mental load and propose original indicators to capture them through self-reported perceptions of organizational responsibility, emotional fatigue, and cognitive spillovers into the workplace.\par
Our findings highlight clear gender asymmetries across all dimensions. Both women and men are more likely to report that, in their couple, the female partner bears primary responsibility for organizing household and childcare activities. Women express lower satisfaction with the division of this responsibility and report higher levels of emotional fatigue. They are also substantially more likely than men to think about the organization of unpaid work while at their paid job, shedding light on the \say{boundarylessness} of cognitive labor. These patterns are consistent across multiple indicators and reflect both the gendered allocation of responsibilities and the invisible nature of mental load. Our results inform the policy debate on how integrating cognitive dimensions into measures of unpaid labor can be crucial to fully capturing the scope and persistence of gender inequality.\par

Our analysis shows that women who report being the primary organizers also spend more time than their partners on household and childcare tasks. In other words, when women spend significantly more time than their partners on household and childcare tasks, they also bear greater organizational responsibilities. In this respect, our study suggests that the gender gap in unpaid work—measured through time use—can serve as a \textit{proxy indicator} of cognitive load. Researchers working with time use data can therefore reasonably infer that larger gender gaps in unpaid work correspond to higher levels of cognitive load borne by women.

Importantly, our dyadic data allow us to compare male and female reports within the same couple, showing that partners frequently diverge in their perceptions of who is responsible for organizing unpaid work. These discrepancies provide compelling descriptive evidence of how mental load is not only unequally distributed, but also unequally recognized. In this respect, our paper complements recent work on the quality of partner relationships within households. \cite{roman2023children} investigate how having children affects couples’ relationship quality. Using a novel measure, they document a sharp and lasting decline in relationship quality immediately after childbirth. They attribute this effect to increased household specialization, as traditional gender-based roles tend to re-emerge after birth, regardless of the pre-existing division of tasks. Our findings suggest that part of this decline may stem from male partners’ lack of awareness of the cognitive and emotional burdens experienced by mothers, which, in turn, may contribute to increased tension within the couple. In future work, we aim to provide a deeper understanding of how mismatched perceptions and expectations may contribute to household conflict and (dis)satisfaction.

Our results indicate that inequality within couples is multidimensional, with interlocking disparities that mutually reinforce one another. Relational asymmetries—ranging from the delegation of organizational tasks to the unequal burden of execution—contribute to women’s emotional exhaustion and may spill over into their professional lives, affecting their labor market outcomes. These cognitive spillovers add a further dimension to the disadvantages women face in the labor market—one that has largely been overlooked. It is well established that employers often anticipate reduced working hours from women after motherhood. Our findings suggest that employers may also perceive mothers as less effective workers, even when they maintain their labor supply, due to the cognitive and emotional burdens associated with family responsibilities. This perception extends the impact of unpaid care work beyond labor supply, shaping further employer expectations and hiring behavior, and thereby possibly reinforcing gender disparities on the demand side of the labor market.

The words of Lord Kelvin—“\textit{When you can measure what you are speaking about, and express it in numbers, you know something about it; but when you cannot measure it, when you cannot express it in numbers, your knowledge is of a meagre and unsatisfactory kind}”—capture why mental load has long been disregarded. What cannot be measured is easily dismissed, even when its burden is deeply felt by those who carry it.\par

Our work gives form and visibility to a hidden burden that disproportionately affects women, whether or not they are in paid employment. Naming and measuring mental load is a necessary first step. Only by recognizing it as real and consequential can we begin to see it as another layer of inequality—often invisible, but deeply felt.

This recognition opens new avenues for policy design, particularly around work-life reconciliation, caregiving support, and gender-sensitive organizational practices. Future research could build on our indicators to assess the effectiveness of such interventions—not only in reducing the time devoted to unpaid work, but also in alleviating its cognitive and emotional load. However, what happens within the home—the division of unpaid work and the sharing of organizational responsibilities—is not easily shaped by policy. Moreover, work-family reconciliation measures do not necessarily help women break free from traditional gender roles as primary caregivers and guardians of family well-being. The only lasting remedy lies in cultural change: dismantling gender norms and enabling couples to share all parenting, domestic, and organizational responsibilities equally.\\

\newpage
\bibliographystyle{chicago}
\bibliography{main}  

\clearpage
\newpage
\appendix
\appendixpage

\section*{Appendix A}
In this Appendix, we report additional estimations not included in the main text and the tables with full estimates presented in the paper.

\begin{table}[H]\centering
\caption{Perception of division of the organization of household activities}
\label{tab:CL_partner}
\resizebox{.9\textwidth}{!}{%
\begin{tabular}{lccccccc}
\toprule
Respondent / Partner & Exclusively & Mostly & Both & Mostly & Exclusively & Total \\
 & Mother & Mother & equally & Father & Father & \\
\midrule
Exclusively mother & 22 & 27 & 14 & 1 & 0 & 64 \\
                   & (5.30) & (6.51) & (3.37) & (0.24) & (0.00) & (15.42) \\
Mostly mother      & 4 & 94 & 60 & 5 & 0 & 163 \\
                   & (0.96) & (22.65) & (14.46) & (1.20) & (0.00) & (39.28) \\
Both equally       & 1 & 25 & 148 & 0 & 0 & 174 \\
                   & (0.24) & (6.02) & (35.66) & (0.00) & (0.00) & (41.93) \\
Mostly father      & 0 & 4 & 4 & 4 & 1 & 13 \\
                   & (0.00) & (0.96) & (0.96) & (0.96) & (0.24) & (3.13) \\
Exclusively father & 0 & 0 & 1 & 0 & 0 & 1 \\
                   & (0.00) & (0.00) & (0.24) & (0.00) & (0.00) & (0.24) \\
\midrule
Total              & 27 & 150 & 227 & 10 & 1 & 415 \\
                   & (6.51) & (36.14) & (54.70) & (2.41) & (0.24) & (100.00) \\
\bottomrule
\end{tabular}
}
\begin{minipage}{\textwidth} 
        \scriptsize \textit{Notes:} Frequencies are shown on the first row of each cell; percentages (of the total sample) are in parentheses below. Columns represent the partner’s responses; rows represent the female respondent’s self-assessment. Diagonal cells indicate agreement between partners on who is primarily responsible for the organization of household activities. The female respondent’s answer is used as the benchmark and compared with her partner’s perception.
    \end{minipage} 
\end{table}

\renewcommand{\arraystretch}{0.9} 
\begin{table}[H]\centering
\caption{Perception of division of the organization of children’s activities}
\label{tab:EL_partner}
\resizebox{.9\textwidth}{!}{%
\begin{tabular}{lccccccc}
\toprule
\textbf{Respondent / Partner} & Exclusively & Mostly & Both & Mostly & Exclusively & Total \\
 & Mother & Mother & equally & Father & Father & \\
\midrule
Exclusively mother & 9 & 7 & 4 & 1 & 0 & 21 \\
                   & (2.17) & (1.69) & (0.96) & (0.24) & (0.00) & (5.06) \\
Mostly mother      & 3 & 99 & 60 & 4 & 0 & 166 \\
                   & (0.72) & (23.86) & (14.46) & (0.96) & (0.00) & (40.00) \\
Both equally       & 0 & 30 & 181 & 8 & 0 & 219 \\
                   & (0.00) & (7.23) & (43.61) & (1.93) & (0.00) & (52.77) \\
Mostly father      & 0 & 1 & 4 & 1 & 1 & 7 \\
                   & (0.00) & (0.24) & (0.96) & (0.24) & (0.24) & (1.69) \\
Exclusively father & 0 & 0 & 2 & 0 & 0 & 2 \\
                   & (0.00) & (0.00) & (0.48) & (0.00) & (0.00) & (0.48) \\
\midrule
Total              & 12 & 137 & 251 & 14 & 1 & 415 \\
                   & (2.89) & (33.01) & (60.48) & (3.37) & (0.24) & (100.00) \\
\bottomrule
\end{tabular}
} 
\begin{minipage}{\textwidth} 
        \scriptsize \textit{Notes:} Frequencies are shown on the first row of each cell; percentages (of the total sample) are in parentheses below. Columns represent the partner’s responses; rows represent the female respondent’s self-assessment. Diagonal cells indicate agreement between partners on who is primarily responsible for the organization of children's activities. The female respondent’s answer is used as the benchmark and compared with her partner’s perception.
    \end{minipage} 
\end{table}

\begin{table}[H]
    \centering
      \caption{Mean difference tests in mental load variables between working and non-working women}
      \resizebox{\textwidth}{!}{%
    \begin{tabular}{lcccccccc}
        \hline
     &   Unemp. &   Mean & Employed &   Mean &   Dif &   St Err &     P-value \\ \hline
Responsible organization HH activities & 123 & 0.740 & 292 & .466 & .274 & .052  & .000 \\ 
Responsible organization children's activities & 123 & 0.708 & 292 & .343 & .365 & .051  & .000 \\ 
 Satisfaction division of organization of HH activities & 123 & 70.512 & 292 & 65.907  & 2.502 & 1.85 & .067 \\ 
 Satisfaction division of organization of children's activities & 123 & 74.529 & 292 & 70.202 & 4.327 & 2.288 & .060 \\ 
 Feeling of fatigue (emotional load) & 123 & 59.541 & 292 & 57.145 & 2.395 & 2.92 & .413 \\ 

\hline
    \end{tabular}
    }
    \label{tab:mean_emp}
\end{table}

\begin{figure}[H]
\centering
\caption{Woman main organizer of HH activities and time allocation - 
working \textit{vs} non-working women}
\label{fig:cl_emp}
{\includegraphics[width=0.9\linewidth]{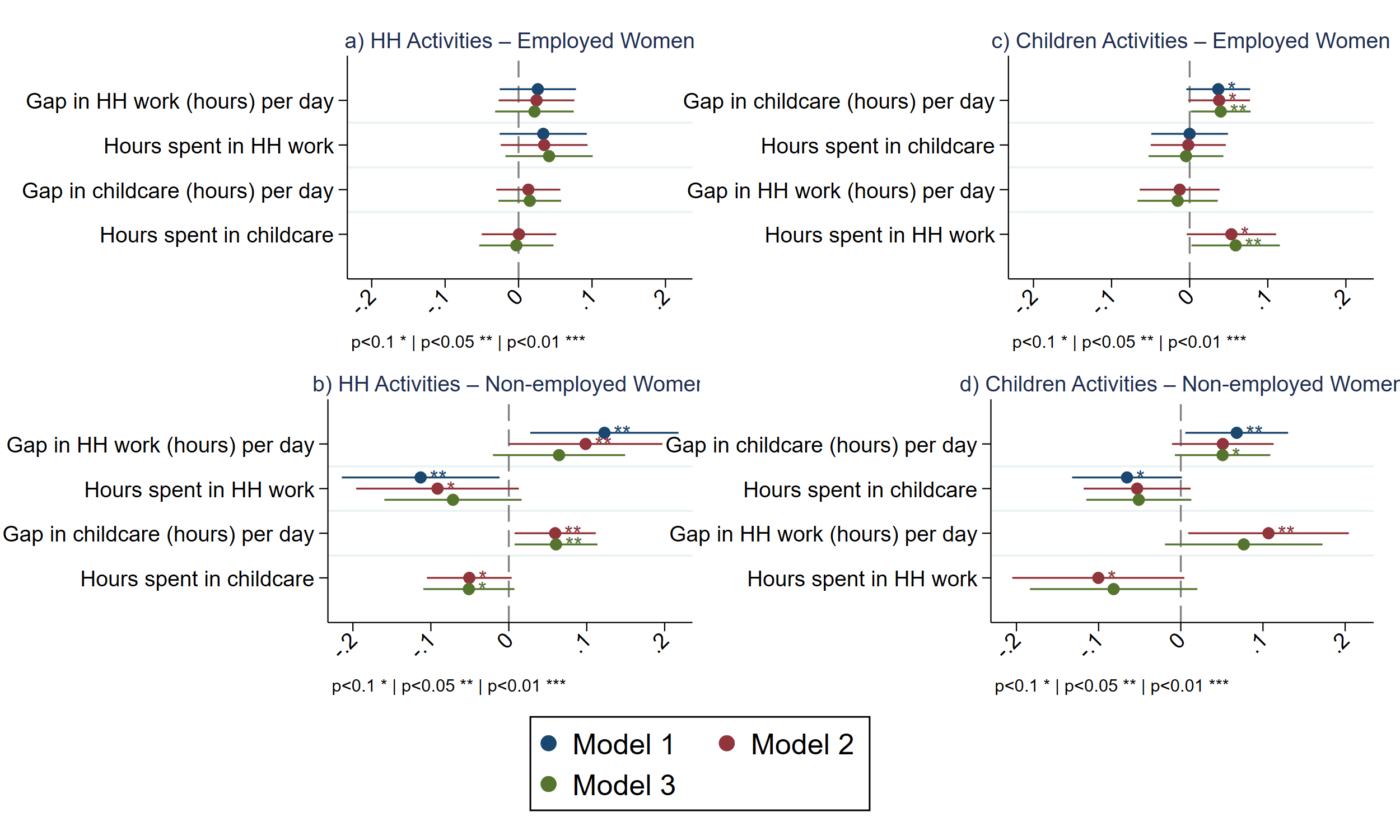}}
\begin{minipage}{\textwidth} 
        \scriptsize \textit{Notes:} The variable of interest is a binary indicator equal to 1 if the respondent reports being the main responsible for the organization of HH activities. In our sample, 56.5\% of unemployed women and 43.5\% of employed women report being the main organizer of HH activities. 
    \end{minipage} 

\end{figure}
\begin{figure}[H]
\centering
\caption{Satisfaction with the organization of HH and children's activities - working vs non-working women}
\label{fig:satisf_emp}
{\includegraphics[width=1\linewidth]{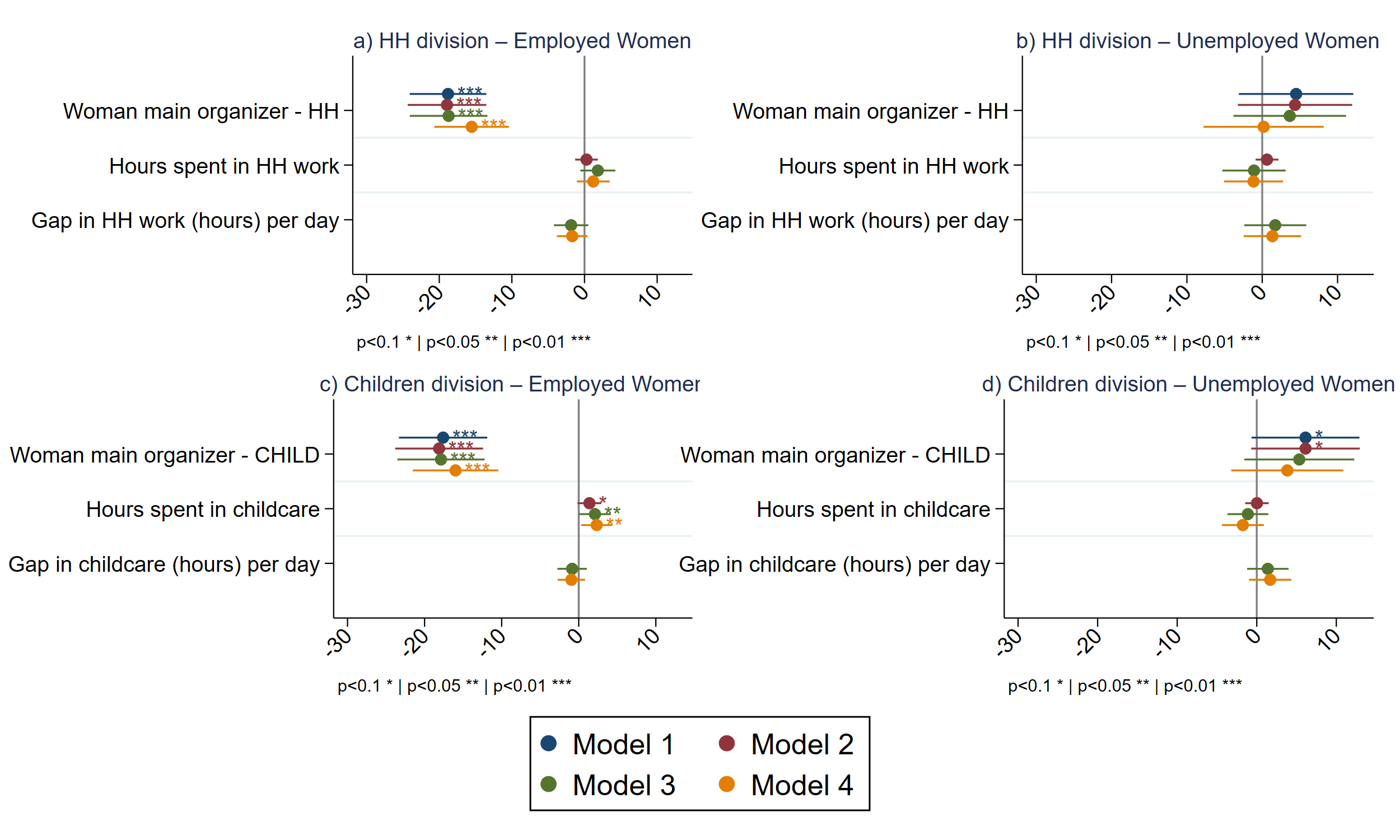}}
\label{fig:10}
\begin{minipage}{\textwidth} 
        \scriptsize \textit{Notes:} *** p$<$0.01; ** p$<$0.05; * p$<$0.10. *** p$<$0.01; ** p$<$0.05; * p$<$0.10. The dependent variable is a satisfaction scale (0–100) for the question: "How satisfied are you with the division of the organization of children's activities?" (employed women mean = 64.06; unemployed women mean = 64.7).
    \end{minipage} 

\end{figure}

\begin{figure}[H]
\centering
\caption{Emotional load - working vs non-working women}
{\includegraphics[width=0.9\linewidth]{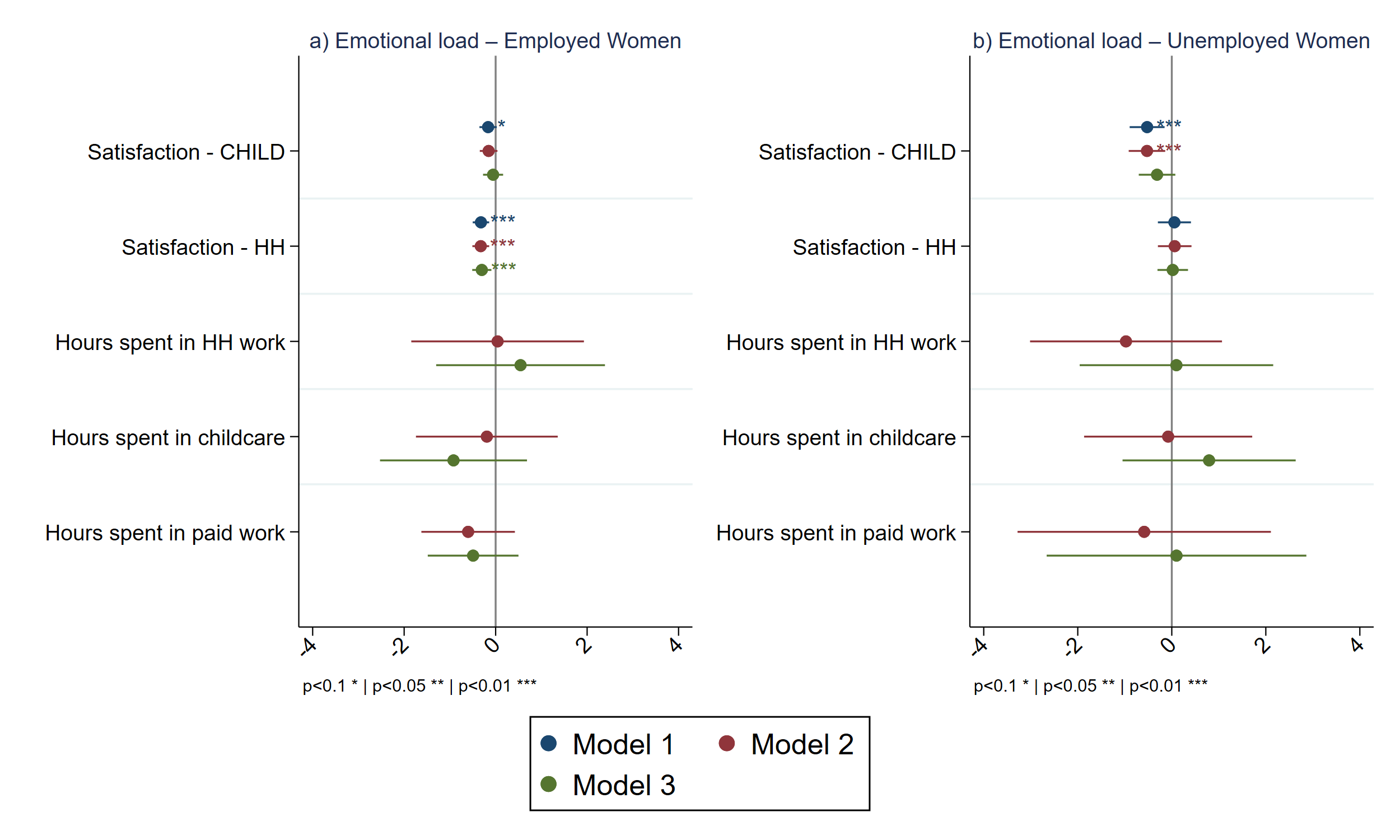}}
\label{fig:fatigue_emp_vs_unemp}
\begin{minipage}{\textwidth} 
       \scriptsize \textit{Notes:} *** p$<$0.01; ** p$<$0.05; * p$<$0.10. *** p$<$0.01; ** p$<$0.05; * p$<$0.10. The feeling of fatigue is measured as the average score of four variables assessing fatigue across different dimensions: children's well-being, partner's well-being, execution of daily activities, and overall family well-being, all measured on a scale from 0 to 100 (employed women mean = 59.2; unemployed women mean = 61.5).
    \end{minipage} 

\end{figure}

\begin{sidewaystable}[ht]
\caption{{\protect\small Gender differences: division of organization of household and children's activities}} 
\label{tab:CL}
\centering
\scriptsize
\begin{threeparttable}
\begin{tabular}{lcccccc|cccccc}
\toprule
& \multicolumn{6}{c|}{\textbf{Women}} & \multicolumn{6}{c}{\textbf{Men}} \\
& (1) & (2) & (3) & (4) & (5) & (6) & (1) & (2) & (3) & (4) & (5) & (6) \\
& \multicolumn{3}{c}{Organization – HH} & \multicolumn{3}{c|}{Organization – Child} 
& \multicolumn{3}{c}{Organization – HH} & \multicolumn{3}{c}{Organization – Child} \\
\cmidrule(lr){2-4} \cmidrule(lr){5-7} \cmidrule(lr){8-10} \cmidrule(lr){11-13}
\midrule
Gap in HH work (hrs)     & 0.054* & 0.046* & 0.035 &       & 0.024 & 0.007   & -0.005 & -0.004 & -0.005 &       & -0.002 & -0.003 \\
                         & (2.47) & (2.10) & (1.53) &       & (1.12) & (0.33) & (1.28) & (1.07) & (1.52) &       & (0.52) & (0.75) \\
Hours in HH work         & 0.004  & 0.008  & 0.005  &       & 0.029  & 0.022  & 0.022  & 0.022  & 0.004  &       & 0.025  & 0.022  \\
                         & (0.17) & (0.33) & (0.20) &       & (1.17) & (0.90) & (1.49) & (1.47) & (0.26) &       & (1.64) & (1.56) \\
Gap in childcare (hrs)   &        & 0.033  & 0.028  & 0.058** & 0.052** & 0.045** &        & -0.007* & -0.007* & -0.011* & -0.009* & -0.010* \\
                         &        & (1.89) & (1.57) & (3.26) & (3.01) & (2.68) &        & (2.26) & (2.26) & (2.57) & (2.45) & (2.45) \\
Hours in childcare       &        & -0.011 & -0.017 & -0.013 & -0.014 & -0.024 &        & -0.007 & -0.007 & -0.009 & -0.010 & -0.011 \\
                         &        & (0.55) & (0.85) & (0.62) & (0.71) & (1.21) &        & (0.88) & (0.87) & (1.14) & (1.25) & (1.35) \\
Employed                 &        &        & -0.208*** &      &       & -0.305*** &       &        & -0.337 &       &        & -0.023 \\
                         &        &        & (3.51) &       &       & (5.13)   &        &        & (1.89) &       &        & (0.22) \\
N of children            &        &        & 0.024  &       &       & 0.020  &        &        & -0.017 &       &        & -0.033** \\
                         &        &        & (0.56) &       &       & (0.49) &        &        & (1.61) &       &        & (2.88) \\
College degree or more  &        &        & 0.058  &       &       & 0.045  &        &        & 0.007  &       &        & -0.007 \\
                         &        &        & (1.06) &       &       & (0.86) &        &        & (0.44) &       &        & (0.37) \\
Constant                 & 0.442*** & 0.442*** & 0.580*** & 0.431*** & 0.324*** & 0.568*** & 0.018 & 0.034 & 0.402* & 0.061** & 0.045 & 0.121 \\
                         & (9.96) & (7.66) & (5.57) & (9.00) & (5.64) & (5.60) & (1.27) & (1.55) & (2.19) & (2.88) & (1.91) & (1.11) \\
Observations             & 415    & 415    & 415    & 415    & 415    & 415    & 415    & 415    & 415    & 415    & 415    & 415 \\
\bottomrule
\end{tabular}
\begin{tablenotes}[para,flushleft]\footnotesize
\item \textit{Note:} *** p$<$0.01; ** p$<$0.05; * p$<$0.10. The dependent variable is a dummy that equals 1 if women reported being the main person organizing household (HH) or children's (Child) activities; for men, it equals 1 if they reported their partner as the main organizer. In the sample, 54.7\% of women report being the main organizer for household activities and 45\% for children's activities, while 42.6\% of men report their partner as the main organizer for household activities and 35.9\% of children's activities. Column (1) includes the household work gap and the time spent in household work; Column (2) adds time spent in childcare and the childcare gap; Column (3) further includes controls for employment status (employed = 1), number of children, and a dummy for holding a university degree or higher.
Column (4) includes the childcare gap and hours spent on childcare. Column (5) adds the HH work gap and hours spent on HH work. Column (6) includes the same controls as in Column (3). The same model specifications apply to the male sample, with dummies indicating whether the respondent’s partner is the main organizer.
\end{tablenotes}
\end{threeparttable}
\end{sidewaystable}

\begin{sidewaystable}[ht]
\caption{{\protect\small Gender differences: satisfaction with division of organization of household and children's activities}} 
\label{tab:satisf}
\centering
\scriptsize
\resizebox{.9\textwidth}{!}{
\begin{threeparttable}
\begin{tabular}{l*{8}{c}|*{8}{c}}
\toprule
& \multicolumn{8}{c|}{\textbf{Women}} & \multicolumn{8}{c}{\textbf{Men}} \\
& (1) & (2) & (3) & (4) & (5) & (6) & (7) & (8) & (1) & (2) & (3) & (4) & (5) & (6) & (7) & (8) \\
& \multicolumn{4}{c}{Satisfaction – organization HH} & \multicolumn{4}{c|}{Satisfaction – organization Child} & \multicolumn{4}{c}{Satisfaction – organization HH} & \multicolumn{4}{c}{Satisfaction – organization Child} \\
\cmidrule(lr){2-5} \cmidrule(lr){6-9} \cmidrule(lr){10-13} \cmidrule(lr){14-17}
\midrule
Main organizer – HH        & -11.18\sym{***} & -11.87\sym{***} & -11.81\sym{***} & -12.01\sym{***} &         &         &         &         & 4.16\sym{*} & 4.71\sym{*} & 4.84\sym{*} & 5.12\sym{*} &         &         &         &         \\
                           & (5.06)          & (5.22)          & (5.26)          & (5.58)          &         &         &         &         & (2.07)      & (2.32)      & (2.25)      & (2.41)      &         &         &         &         \\
Hours in HH work           &                 & 0.95            & 1.20            & 0.62            &         &         &         &         &             & 1.35        & 1.22        & 0.13        &         &         &         &         \\
                           &                 & (1.62)          & (1.04)          & (0.61)          &         &         &         &         &             & (1.39)      & (1.20)      & (0.13)      &         &         &         &         \\
Gap in HH work (hrs)       &                 &                 & -0.26           & -0.88           &         &         &         &         &             &             & -0.17       & -0.44       &         &         &         &         \\
                           &                 &                 & (0.24)          & (0.90)          &         &         &         &         &             &             & (0.28)      & (0.74)      &         &         &         &         \\
                           Main organizer – CHILD     &                 &                 &                 &                 & -8.42\sym{***} & -8.94\sym{***} & -8.99\sym{***} & -10.51\sym{***} &             &             &             &             & 0.83     & 1.37     & -0.21    & 0.82  \\
                           &                 &                 &                 &                 & (3.95)         & (4.18)         & (4.21)         & (4.62)          &             &             &             &             & (0.44)   & (0.72)   & (0.11)  & (0.44) \\
Hours in childcare         &                 &                 &                 &                 &         & 1.06     & 0.95     & 0.70     &             &             &             &             &         & 2.16\sym{**} & 3.71\sym{***} & 3.44\sym{***} \\
                           &                 &                 &                 &                 &         & (1.85)   & (1.07)   & (0.81)   &             &             &             &             &         & (3.26)   & (5.07) & (4.84) \\
Gap in childcare (hrs)     &                 &                 &                 &                 &         &         & 0.12     & -0.06    &             &             &             &             &         &         & 2.16\sym{***} & 2.06\sym{***} \\
                           &                 &                 &                 &                 &         &         & (0.14)   & (0.07)   &             &             &             &             &         &         & (4.31) & (4.33) \\
Employed                   &                 &                 &                 & -4.08           &         &         &         & -5.02   &             &             &             & -14.51\sym{**} &         &         &         & -14.33\sym{***} \\
                           &                 &                 &                 & (1.61)          &         &         &         & (1.95)  &             &             &             & (2.62)      &         &         &         & (3.41) \\
\# of children              &                 &                 &                 & -2.40           &         &         &         & -6.56\sym{**} &             &             &             & -6.57\sym{***} &         &         &         & -6.62\sym{***} \\
                           &                 &                 &                 & (1.11)          &         &         &         & (3.03)  &             &             &             & (4.06)      &         &         &         & (4.26) \\
College degree or more     &                 &                 &                 & -16.03\sym{***} &         &         &         & -9.87\sym{***} &             &             &             & -2.99       &         &         &         & -2.42 \\
                           &                 &                 &                 & (6.93)          &         &         &         & (4.52)  &             &             &             & (1.46)      &         &         &         & (1.33) \\
Constant                   & 73.39\sym{***}  & 71.34\sym{***}  & 71.14\sym{***}  & 86.27\sym{***}  & 75.28\sym{***}  & 72.88\sym{***}  & 73.06\sym{***}  & 91.00\sym{***} & 73.78\sym{***} & 72.45\sym{***} & 72.80\sym{***} & 98.56\sym{***} & 77.11\sym{***} & 73.48\sym{***} & 69.64\sym{***} & 93.99\sym{***} \\
                           & (48.24)         & (36.77)         & (35.63)         & (21.06)         & (63.96)         & (42.08)         & (36.65)         & (23.06)         & (54.65)     & (44.61)     & (38.77)     & (15.44)     & (67.15)     & (45.63)     & (38.75)     & (19.38) \\
Observations               & 415             & 415             & 415             & 415             & 415             & 415             & 415             & 415             & 415         & 415         & 415         & 415         & 415         & 415         & 415         & 415 \\
\bottomrule
\end{tabular}
\begin{tablenotes}[para,flushleft]\footnotesize
\item \textit{Note:} *** p$<$0.01; ** p$<$0.05; * p$<$0.10. The dependent variable is a satisfaction scale (0–100) for the question: *"How satisfied are you with the division of the organization of HH activities?"* (Women mean = 67.27; Men mean = 75.55) and ”How satisfied are you with the division of the organization of children’s activities?”
(women mean = 71.48; men mean = 77.40). For women: Column (1) includes a dummy equal to 1 if the respondent is the main organizer of HH activities. Column (2) adds the HH work gap and the respondent’s hours spent on HH work. Column (3) adds the childcare gap and hours spent on childcare. Column (4) further includes controls for employment status (employed = 1), number of children, and a dummy for having a university degree or higher. Column (5) includes a dummy for being the main organizer of children’s activities. Column (6) adds the childcare gap and hours spent on childcare. Column (7) adds the HH work gap and hours spent on HH work. Column (8) includes the same controls as in Column (4). The same model specifications apply to the male sample, with dummies indicating whether the respondent’s partner is the main organizer.
\end{tablenotes}

\end{threeparttable}
}
\end{sidewaystable}

\begin{sidewaystable}[ht]
\caption{{\protect\small Gender differences: thinking about the organization of household and children’s activities at work}} 
\label{tab:think}
\centering
\scriptsize
\resizebox{.9\textwidth}{!}{
\begin{threeparttable}
\begin{tabular}{l*{8}{c}|*{8}{c}}
\toprule
& \multicolumn{8}{c|}{\textbf{Women}} & \multicolumn{8}{c}{\textbf{Men}} \\
& (1) & (2) & (3) & (4) & (5) & (6) & (7) & (8) 
& (1) & (2) & (3) & (4) & (5) & (6) & (7) & (8) \\
& \multicolumn{4}{c}{Think – HH} & \multicolumn{4}{c|}{Think – CHILD} 
& \multicolumn{4}{c}{Think – HH} & \multicolumn{4}{c}{Think – CHILD} \\
\cmidrule(lr){2-5} \cmidrule(lr){6-9} \cmidrule(lr){10-13} \cmidrule(lr){14-17}
\midrule
Main organizer – HH              
& 0.222\sym{***} & 0.216\sym{***} & 0.214\sym{***} & 0.242\sym{***} &       &       &       &       %
& -0.0387 & -0.0210 & -0.0190 & -0.0176 &       &       &       &       \\
& (3.91)  & (3.77)  & (3.71)  & (4.21)  &       &       &       &       %
& (1.35)  & (0.71)  & (0.62)  & (0.57)  &       &       &       &       \\
Hours spent in HH work          
&         & 0.0032  & -0.0128 & -0.0190 &       &       &       &       %
&         & 0.0586\sym{**} & 0.0565\sym{**} & 0.0574\sym{**} &       &       &       &       \\
&         & (0.16)  & (0.48)  & (0.71)  &       &       &       &       %
&         & (2.86)  & (2.78)  & (2.79)  &       &       &       &       \\
Hours spent in paid work        
&         & -0.0110 & -0.0106 & -0.0112 &       & 0.0033 & 0.0035 & 0.0032 %
&         & 0.0101\sym{*} & 0.0102\sym{*} & 0.0105\sym{*} &       & 0.0055 & 0.0059 & 0.0060 \\
&         & (1.13)  & (1.08)  & (1.16)  &       & (0.33) & (0.36) & (0.33) %
&         & (2.26)  & (2.23)  & (2.30)  &       & (1.01) & (1.07) & (1.07) \\
Gap in HH work (hrs/day)       
&         &         & 0.0190  & 0.0197  &       &       &       &       %
&         &         & -0.0026 & -0.0026 &       &       &       &       \\
&         &         & (0.82)  & (0.85)  &       &       &       &       %
&         &         & (0.34)  & (0.32)  &       &       &       &       \\
Main organizer – CHILD          
&         &         &         &         & 0.300\sym{***} & 0.285\sym{***} & 0.283\sym{***} & 0.298\sym{***} %
&         &         &         &         & -0.0438 & -0.0395 & -0.0353 & -0.0346 \\
&         &         &         &         & (5.11) & (4.84) & (4.77) & (5.01) %
&         &         &         &         & (1.29) & (1.17) & (1.00) & (0.98) \\
Hours spent in childcare        
&         &         &         &         &        & 0.0453\sym{**} & 0.0381 & 0.0402 %
&         &         &         &         &        & 0.0218 & 0.0176 & 0.0171 \\
&         &         &         &         &        & (2.90) & (1.67) & (1.75) %
&         &         &         &         &        & (1.49) & (1.15) & (1.10) \\
Gap in childcare (hrs/day)      
&         &         &         &         &        &        & 0.0087 & 0.0084 %
&         &         &         &         &        &        & -0.0059 & -0.0064 \\
&         &         &         &         &        &        & (0.45) & (0.43) %
&         &         &         &         &        &        & (0.79) & (0.83) \\
Total number of children        
&         &         &         & -0.0621 &       &       &       & -0.0905\sym{*} %
&         &         &         & 0.0146  &       &       &       & -0.0019 \\
&         &         &         & (1.14)  &       &       &       & (1.98) %
&         &         &         & (0.47)  &       &       &       & (0.06) \\
College degree or more          
&         &         &         & -0.115\sym{*} &   &   &   & -0.0665 %
&         &         &         & -0.0331 &       &       &       & -0.0150 \\
&         &         &         & (2.00)  &       &       &       & (1.18) %
&         &         &         & (1.10)  &       &       &       & (0.42) \\
Constant                         
& 0.308\sym{***} & 0.351\sym{***} & 0.361\sym{***} & 0.502\sym{***} & 0.370\sym{***} & 0.261\sym{***} & 0.272\sym{***} & 0.419\sym{***} %
& 0.113\sym{***} & 0.0030 & 0.0075 & -0.0027 & 0.151\sym{***} & 0.0837 & 0.0916\sym{*} & 0.101 \\
& (8.30) & (4.74) & (4.81) & (4.66) & (10.58) & (3.77) & (3.72) & (4.24) %
& (5.40) & (0.09) & (0.23) & (0.05) & (6.76) & (1.93) & (2.12) & (1.69) \\
Observations                     
& 292    & 292    & 292    & 292    & 292    & 292    & 292    & 292 %
& 407    & 407    & 407    & 407    & 407    & 407    & 407    & 407 \\
\bottomrule
\end{tabular}
\begin{tablenotes}[para,flushleft]\footnotesize
\item \textit{Note:} *** p$<$0.01; ** p$<$0.05; * p$<$0.10. 
The models are estimated on respondents who self-report to have a job (70.3\% (N=292/415) working women and 98.07\% (N=408/415) working men.) The dependent variable is a binary indicator equal to 1 if the respondent reports thinking “very often” or “always” about the organization of household or children’s activities while at work. For women: Column (1) includes a dummy equal to 1 if the respondent is the main organizer of HH activities. Column (2) adds the HH work gap and the respondent’s hours spent on HH work. Column (3) adds the childcare gap and hours spent on childcare. Column (4) further includes controls for the number of children, and a dummy for having a university degree or higher. Column (5) includes a dummy for being the main organizer of children’s activities. Column (6) adds the childcare gap and hours spent on childcare. Column (7) adds the HH work gap and hours spent on HH work. Column (8) includes the same controls as in Column (4). The same model specifications apply to the male sample, with dummies indicating whether the respondent’s partner is the main organizer.
\end{tablenotes}
\end{threeparttable}
}
\end{sidewaystable}

\begin{table}[ht]
\caption{{\protect\small Gender differences: feeling of fatigue}}
\label{tab:fatigue}
\centering
\scriptsize
\begin{threeparttable}
\begin{tabular}{lccc|ccc}
\toprule
& \multicolumn{3}{c|}{\textbf{Women}} & \multicolumn{3}{c}{\textbf{Men}} \\
& (1) & (2) & (3) & (1) & (2) & (3) \\
& \multicolumn{3}{c|}{Mean feeling of fatigue} & \multicolumn{3}{c}{Mean feeling of fatigue} \\
\midrule
Satisfaction – CHILD          
& -0.237\sym{**} & -0.230\sym{**} & -0.112         & -0.455\sym{***} & -0.439\sym{***} & -0.355\sym{**} \\
& (2.70)         & (2.60)         & (1.15)         & (3.84)         & (3.66)         & (2.81)         \\
Satisfaction – HH             
& -0.232\sym{**} & -0.238\sym{**} & -0.232\sym{*}  & -0.070         & -0.078         & -0.040         \\
& (2.80)         & (2.83)         & (2.59)         & (0.67)         & (0.73)         & (0.34)         \\
Hours spent in HH work        
&                & -0.266         & 0.293          &                & 0.870          & 1.116          \\
&                & (0.39)         & (0.43)         &                & (0.61)         & (0.80)         \\
Hours spent in childcare       
&                & -0.234         & -0.457         &                & -0.846         & -0.641         \\
&                & (0.40)         & (0.77)         &                & (0.84)         & (0.65)         \\
Hours spent in paid work       
&                & -0.805         & -0.495         &                & -0.463         & -0.392         \\
&                & (1.80)         & (1.03)         &                & (1.05)         & (0.93)         \\
Unemployed                    
&                &               & 0              &                &               & 0              \\
&                &               & (.)            &                &               & (.)            \\
Full time                      
&                &               & -4.066         &                &               & 7.670          \\
&                &               & (1.23)         &                &               & (1.50)         \\
Part time                     
&                &               & -0.409         &                &               & 23.38\sym{**}  \\
&                &               & (0.10)         &                &               & (2.98)         \\
Total Number of Children       
&                &               & 12.99\sym{***} &                &               & 14.39\sym{***} \\
&                &               & (7.82)         &                &               & (6.99)         \\
College Degree or more         
&                &               & 7.303\sym{**}  &                &               & 0.665          \\
&                &               & (2.98)         &                &               & (0.26)         \\
Constant                       
& 90.44\sym{***} & 94.18\sym{***} & 64.83\sym{***} & 93.71\sym{***} & 96.29\sym{***} & 58.02\sym{***} \\
& (23.89)        & (19.42)        & (9.64)         & (15.33)        & (14.16)        & (5.88)         \\
\midrule
Observations                   
& 415            & 415            & 415            & 415            & 415            & 415            \\
\bottomrule
\end{tabular}
\begin{tablenotes}[para,flushleft]\footnotesize
\item \textit{Note:} *** p$<$0.01; ** p$<$0.05; * p$<$0.10. The feeling of fatigue is measured as the mean of four variables measuring, on a scale from 0 to 100, the level of fatigue for four dimensions: children's well-being; partner's well-being; execution of daily activities; and overall family's well-being (women mean = 60.3; men mean = 57). Column (1) includes the levels of satisfaction with the division of household and children's activities. Column (2) adds the HH work gap, the respondent’s hours spent on HH work, the childcare gap and hours spent on childcare. Column (3) further includes controls for employment status (employed = 1), number of children, and a dummy for having a university degree or higher.  The same model specifications apply to the male sample.
\end{tablenotes}
\end{threeparttable}
\end{table}

\begin{sidewaystable}[ht]
\caption{{\protect\small Organization of household and children's activities: employed vs. unemployed women}} 
\label{tab:CL_emp}
\centering
\scriptsize
\begin{threeparttable}
\begin{tabular}{l*{6}{c}|*{6}{c}}
\toprule
& \multicolumn{6}{c|}{\textbf{Employed Women}} & \multicolumn{6}{c}{\textbf{Unemployed Women}} \\
& (1) & (2) & (3) & (4) & (5) & (6) & (1) & (2) & (3) & (4) & (5) & (6) \\
& \multicolumn{3}{c}{Organization – HH} & \multicolumn{3}{c|}{Organization – Child} 
& \multicolumn{3}{c}{Organization – HH} & \multicolumn{3}{c}{Organization – Child} \\
\cmidrule(lr){2-4} \cmidrule(lr){5-7} \cmidrule(lr){8-10} \cmidrule(lr){11-13}
\midrule
Gap in HH work (hrs/day)
& 0.026 & 0.024 & 0.022 &       & -0.013 & -0.015
& 0.123\sym{*} & 0.099\sym{*} & 0.065 &       & 0.107\sym{*} & 0.077 \\
& (0.99) & (0.93) & (0.80) &       & (0.49) & (0.59)
& (2.55) & (1.98) & (1.51) &       & (2.16) & (1.58) \\

Hours in HH work
& 0.034 & 0.035 & 0.042 &       & 0.054 & 0.059\sym{*}
& -0.113\sym{*} & -0.092 & -0.072 &       & -0.100 & -0.082 \\
& (1.11) & (1.16) & (1.38) &       & (1.84) & (2.05)
& (-2.21) & (-1.74) & (-1.61) &       & (-1.90) & (-1.59) \\

Gap in childcare (hrs/day)
&       & 0.013 & 0.015 & 0.037 & 0.038 & 0.040\sym{*}
&       & 0.060\sym{*} & 0.061\sym{*} & 0.068\sym{*} & 0.051 & 0.051 \\
&       & (0.60) & (0.70) & (1.76) & (1.88) & (2.05)
&       & (2.26) & (2.26) & (2.15) & (1.64) & (1.74) \\

Hours in childcare
&       & 0.001 & -0.003 & -0.000 & -0.002 & -0.005
&       & -0.051 & -0.051 & -0.066 & -0.053 & -0.051 \\
&       & (0.02) & (-0.12) & (-0.00) & (-0.07) & (-0.19)
&       & (-1.85) & (-1.73) & (-1.94) & (-1.62) & (-1.59) \\

Employed
&       &       & 0       &       &       & 0
&       &       & 0       &       &       & 0 \\
&       &       & (.)     &       &       & (.)
&       &       & (.)     &       &       & (.) \\

Total Number of Children
&       &       & 0.033   &       &       & 0.008
&       &       & -0.006  &       &       & 0.025 \\
&       &       & (0.56)  &       &       & (0.15)
&       &       & (-0.08) &       &       & (0.36) \\

College Degree or more
&       &       & 0.184\sym{**} &       &       & 0.165\sym{**}
&       &       & -0.428\sym{***} &       &       & -0.392\sym{***} \\
&       &       & (3.13) &       &       & (2.91)
&       &       & (-3.91) &       &       & (-3.44) \\

Constant
& 0.361\sym{***} & 0.352\sym{***} & 0.215\sym{*} & 0.322\sym{***} & 0.227\sym{***} & 0.134
& 0.782\sym{***} & 0.835\sym{***} & 0.955\sym{***} & 0.799\sym{***} & 0.833\sym{***} & 0.891\sym{***} \\
& (7.00) & (5.20) & (2.08) & (5.93) & (3.53) & (1.42)
& (9.99) & (7.79) & (6.47) & (10.08) & (7.78) & (5.96) \\

Observations
& 292 & 292 & 292 & 292 & 292 & 292
& 123 & 123 & 123 & 123 & 123 & 123 \\
\bottomrule
\end{tabular}
\begin{tablenotes}[para,flushleft]\footnotesize
\item \textit{Note:} *** p$<$0.01; ** p$<$0.05; * p$<$0.10. The dependent variable is a dummy that equals 1 if women reported being the main person organizing children's activities. In the sample, 46.58\% of employed women and 73.98\% of unemployed women report this, respectively. Column (1) includes the household work gap and the time spent in household work; Column (2) adds time spent in childcare and the childcare gap; Column (3) further includes controls for employment status (employed = 1), number of children, and a dummy for holding a university degree or higher. Column (4) includes the childcare gap and hours spent on childcare. Column (5) adds the HH work gap and hours spent on HH work. Column (6) includes the same controls as in Column (3). The same model specifications apply to the unemployed women sample.
\end{tablenotes}
\end{threeparttable}
\end{sidewaystable}

\begin{sidewaystable}[ht]
\caption{{\protect\small Satisfaction with division of organization of household and childcare activities: employed \textit{vs.} unemployed women}}
\label{tab:satisf_emp}
\centering
\scriptsize
\resizebox{.9\textwidth}{!}{
\begin{threeparttable}
\begin{tabular}{l*{8}{c}|*{8}{c}}
\toprule
& \multicolumn{8}{c|}{\textbf{Employed Women}} & \multicolumn{8}{c}{\textbf{Unemployed Women}} \\
& (1) & (2) & (3) & (4) & (5) & (6) & (7) & (8)
& (1) & (2) & (3) & (4) & (5) & (6) & (7) & (8) \\
& \multicolumn{4}{c}{Satisfaction – HH} & \multicolumn{4}{c|}{Satisfaction – Child}
& \multicolumn{4}{c}{Satisfaction – HH} & \multicolumn{4}{c}{Satisfaction – Child} \\
\cmidrule(lr){2-5} \cmidrule(lr){6-9} \cmidrule(lr){10-13} \cmidrule(lr){14-17}
\midrule
Main Organizer HH & -18.79\sym{***} & -18.93\sym{***} & -18.71\sym{***} & -15.53\sym{***} & & & & 
   & 4.494 & 4.349 & 3.646 & 0.187 & & & & \\
   & (7.02) & (6.88) & (6.89) & (5.96) & & & & 
   & (1.17) & (1.14) & (0.97) & (0.05) & & & & \\

Hours in HH work & & 0.274 & 1.837 & 1.205 & & & & 
                & & 0.633 & -1.093 & -1.159 & & & & \\
                & & (0.34) & (1.49) & (1.05) & & & & 
                & & (0.83) & (-0.51) & (-0.59) & & & & \\

Gap HH work hrs & & & -1.841 & -1.690 & & & & 
               & & & 1.722 & 1.349 & & & & \\
               & & & (1.53) & (1.57) & & & & 
               & & & (0.83) & (0.70) & & & & \\

Main organizer Child & & & & & -17.60\sym{***} & -18.12\sym{***} & -17.87\sym{***} & -15.99\sym{***}
      & & & & & 6.127 & 6.132 & 5.342 & 3.838 \\
      & & & & & (6.05) & (6.26) & (6.22) & (5.67)
      & & & & & (1.79) & (1.78) & (1.53) & (1.08) \\

Hours childcare & & & & & & 1.387 & 2.106\sym{*} & 2.336\sym{*}
                & & & & & & 0.023 & -1.119 & -1.746 \\
                & & & & & & (1.77) & (2.01) & (2.26)
                & & & & & & (0.03) & (0.86) & (1.31) \\

Gap childcare hrs & & & & & & & -0.855 & -0.950 
                  & & & & & & & 1.390 & 1.683 \\
                  & & & & & & & (0.88) & (1.05)
                  & & & & & & & (1.06) & (1.25) \\
Total children & & & & -1.679 & & & & -6.769\sym{*} 
               & & & & -3.628 & & & & -7.666\sym{*} \\
               & & & & (0.63) & & & & (2.59) 
               & & & & (1.08) & & & & (2.21) \\

College degree & & & & -15.47\sym{***} & & & & -9.025\sym{***} 
               & & & & -9.213 & & & & -3.675 \\
               & & & & (5.79) & & & & (3.65) 
               & & & & (1.69) & & & & (0.78) \\

Constant & 74.66\sym{***} & 74.14\sym{***} & 73.00\sym{***} & 82.34\sym{***} & 76.23\sym{***} & 73.37\sym{***} & 72.18\sym{***} & 84.63\sym{***}
         & 67.19\sym{***} & 65.07\sym{***} & 66.67\sym{***} & 77.44\sym{***} & 70.19\sym{***} & 70.12\sym{***} & 71.94\sym{***} & 85.99\sym{***} \\
         & (43.89) & (34.24) & (33.70) & (20.38) & (59.40) & (35.55) & (31.44) & (21.22)
         & (21.33) & (16.33) & (16.54) & (11.05) & (24.75) & (18.41) & (18.05) & (13.57) \\
\midrule
Observations & 292 & 292 & 292 & 292 & 292 & 292 & 292 & 292 
             & 123 & 123 & 123 & 123 & 123 & 123 & 123 & 123 \\
\bottomrule
\end{tabular}
\begin{tablenotes}[para,flushleft]\footnotesize
\item \textit{Note:} *** p$<$0.01; ** p$<$0.05; * p$<$0.10. The dependent variable is a satisfaction scale (0–100) for the question: *"How satisfied are you with the division of the organization of HH activities?"* (Employed women mean = 71.35; Unemployed women mean = 71.74) and ”How satisfied are you with the division of the organization of children’s activities?”
(employed women mean = 74.22; unemployed women mean = 75.61). Column (1) includes a dummy equal to 1 if the respondent is the main organizer of HH activities. Column (2) adds the HH work gap and the respondent’s hours spent on HH work. Column (3) adds the childcare gap and hours spent on childcare. Column (4) further includes controls for employment status (employed = 1), number of children, and a dummy for having a university degree or higher. Column (5) includes a dummy for being the main organizer of children’s activities. Column (6) adds the childcare gap and hours spent on childcare. Column (7) adds the HH work gap and hours spent on HH work. Column (8) includes the same controls as in Column (4). The same model specifications apply to the unemployed women sample.
\end{tablenotes}
\end{threeparttable}
}
\end{sidewaystable}

\begin{table}[ht]
\caption{{\protect\small Feeling of fatigue: employed vs. unemployed women}} 
\label{tab:fatigue_emp}
\centering
\scriptsize
\begin{threeparttable}
\begin{tabular}{lccc|ccc}
\toprule
& \multicolumn{3}{c|}{\textbf{Employed Women}} & \multicolumn{3}{c}{\textbf{Unemployed Women}} \\
& (1) & (2) & (3) & (1) & (2) & (3) \\
& \multicolumn{3}{c|}{Mean feeling of fatigue} & \multicolumn{3}{c}{Mean feeling of fatigue} \\
\hline
Satisfaction – Child 
& -0.164 & -0.153 & -0.056 
& -0.525\sym{**} & -0.527\sym{**} & -0.313 \\
& (1.71) & (1.56) & (0.50) 
& (2.80) & (2.69) & (1.59) \\
[1em]
Satisfaction – HH 
& -0.321\sym{***} & -0.325\sym{***} & -0.302\sym{**} 
& 0.056 & 0.063 & 0.021 \\
& (3.51) & (3.49) & (2.84) 
& (0.31) & (0.35) & (0.13) \\
[1em]
Hours in HH work 
&        & 0.043 & 0.545 
&        & -0.973 & 0.099 \\
&        & (0.05) & (0.58) 
&        & (0.94) & (0.10) \\
[1em]
Hours in childcare 
&        & -0.192 & -0.921 
&        & -0.077 & 0.794 \\
&        & (0.24) & (1.13) 
&        & (0.09) & (0.85) \\
[1em]
Hours in paid work 
&        & -0.601 & -0.492 
&        & -0.586 & 0.101 \\
&        & (1.16) & (0.98) 
&        & (0.43) & (0.07) \\
[1em]
Part time 
&        &        & 4.178 
&        &        &       \\
&        &        & (0.47) 
&        &        &       \\
[1em]
Total number of children 
&        &        & 11.64\sym{***} 
&        &        & 16.54\sym{***} \\
&        &        & (5.52) 
&        &        & (5.41) \\
[1em]
College degree or more 
&        &        & 7.151\sym{*} 
&        &        & 7.724 \\
&        &        & (2.36) 
&        &        & (1.98) \\
[1em]
Constant 
& 89.83\sym{***} & 92.21\sym{***} & 63.70\sym{***} 
& 94.71\sym{***} & 98.38\sym{***} & 53.71\sym{***} \\
& (21.55) & (15.63) & (5.58) 
& (10.99) & (10.90) & (4.11) \\
\midrule
Observations 
& 292 & 292 & 292 
& 123 & 123 & 123 \\
\bottomrule
\end{tabular}
\begin{tablenotes}[para,flushleft]\footnotesize
\item \textit{Note:} *** p$<$0.01; ** p$<$0.05; * p$<$0.10. The feeling of fatigue is measured as the mean of four variables measuring, on a scale from 0 to 100, the level of fatigue for four dimensions: children's well-being; partner's well-being; execution of daily activities; and overall family's well-being (employed women mean = 54.93; unemployed women mean = 58.77). Column (1) includes the levels of satisfaction with the division of household and children's activities. Column (2) adds the HH work gap, the respondent’s hours spent on HH work, the childcare gap, and hours spent on childcare. Column (3) further includes controls for employment status (employed = 1), number of children, and a dummy for having a university degree or higher.  The same model specifications apply to the unemployed women sample.
\end{tablenotes}
\end{threeparttable}
\end{table}

\clearpage
\newpage
\section*{Appendix B}
In this section we report an English translation of the exact questions used to measure Mental Load taken from the socio-economic questionnaire filled by the participants. The questionnaire was administered in Italian using a CAWI methodology.

\section*{Measuring Mental Load}

The following questions concern the organization of family life. The organizational work of daily family life includes both the planning of domestic tasks (e.g., thinking about what to cook for dinner, what to buy at the supermarket, when to do the laundry, etc.) and the management of children's activities (e.g., remembering doctor’s appointments, organizing their social life, enrolling them in summer camps, etc.). Note that “organizing” differs from “executing” domestic and family tasks, and that these roles can be carried out by different people.\\

\subsection*{Questions}
Question 1-5 referred to socio-demographic and household composition.\\
\bigskip

\noindent
\textbf{Question 6:} Thinking about your household (Select only one answer per row)\\

\begin{table}[h!]
    \centering
   \renewcommand{\arraystretch}{0.90}
\small
\hspace*{-1cm}
 \begin{tabular}{lccccc}
\hline
& Exclusively & Mostly & Shared & Mostly my & Exclusively \\
&  me &  me & equally & partner & my partner \\
\hline
Who is responsible for organizing & [ ] & [ ] & [ ] & [ ] & [ ] \\
household tasks? &  &  & & & \\
\hline
Who is responsible for organizing & [ ] & [ ] & [ ] & [ ] & [ ]\\
activities related to co-residing children? &  &  & & & \\
\hline
\end{tabular}
\end{table}

\bigskip
\noindent
\textbf{Question 7:} \textit{(Asked only to those who are employed)} Considering the family organization tasks you usually perform, how often do you think about them during a typical workday? (Select only one answer per row)\\

\begin{table}[H]
    \centering
    \hspace*{-1cm}
\begin{tabular}{lccccc}
& Never & Occasionally & Often & Very often & Constantly \\
\hline
Organization of household tasks & [ ] & [ ] & [ ] & [ ] & [ ] \\
\hline
Organization of childcare activities & [ ] & [ ] & [ ] & [ ] & [ ]\\
\hline
\end{tabular}
\end{table}

\bigskip

\noindent
From this point on, for some questions, you will see a slider like this: 
\begin{figure}[H]
    \centering
    \includegraphics[width=0.5\linewidth]{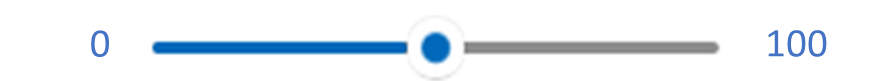}
    \caption{}
    \label{fig:enter-label}
\end{figure}

\noindent Please use the slider to indicate a value between 0 and 100, according to your answer.

\bigskip
\noindent
\textbf{Question 8:} How satisfied are you with the following? (0 = Not at all, 100 = Completely)
\begin{itemize}
    \item Division of responsibility for organizing household tasks
    \item Division of responsibility for organizing childcare activities
\end{itemize}

\bigskip
\noindent
\textbf{Question 9:} Regardless of how many and which organizing tasks you are responsible for, how responsible do you feel for the following? (0 = Not at all, 100 = Completely)
\begin{itemize}
    \item Well-being of co-residing children
    \item Well-being of your partner
    \item Execution of daily life tasks
    \item Maintaining balance within the family
\end{itemize}

\bigskip
\noindent
\textbf{Question 10:} For each area of responsibility listed below, indicate how much it causes you fatigue and/or stress (0 = Not at all, 100 = Completely)
\begin{itemize}
    \item Well-being of co-residing children
    \item Well-being of your partner
    \item Execution of daily life tasks
    \item Maintaining balance within the family
\end{itemize}
\end{document}